\documentclass[%
nofootinbib,
amsmath,amssymb,
aps,showkeys,showpacs,
prd,
]{revtex4-1}

\usepackage[caption=false]{subfig}
\usepackage{amsmath}
\usepackage{xcolor}
\usepackage{hyperref}
\usepackage{graphicx}
\begin{document}

\title[Crowd avoidance model]{Modelling competition for space: Emergent inefficiency and inequality due to spatial self-organization among a group of crowd-avoiding agents}

\author{Ann Mary Mathew}
\email{annmaryettolil@gmail.com}
\affiliation{Department of Physics, Cochin University of Science and Technology}
\affiliation{Assumption College Autonomous, Changanassery}

\author{V. Sasidevan}
\email{sasidevan@gmail.com}
\affiliation{Department of Physics, Cochin University of Science and Technology}

\begin{abstract}
Competition for a limited resource is the hallmark of many complex systems, and often, that resource turns out to be the physical space itself. In this work, we study a novel model designed to elucidate the dynamics and emergence in complex adaptive systems in which agents compete for some spatially spread resource. Specifically, in the model, the dynamics result from the agents trying to position themselves in the quest to avoid physical crowding experienced locally. We characterize in detail the dependence of the emergent behavior of the model on the population density of the system and the individual-level agent traits such as the extent of space an agent considers as her neighborhood, the limit of occupation density one tolerates within that neighborhood, and the information accessibility of the agents about neighborhood occupancy. We show that inefficiency in utilizing physical space shows transition at a specific density and peaks at another distinct density. Furthermore, we demonstrate that the variation of inefficiency relative to the information accessible to the agents exhibits contrasting behavior above and below this second density. We also look into the inequality of resource sharing in the model and show that although inefficiency can be a non-monotonic function of information depending upon the parameters of the model, inequality, in general, decreases with information. Our study sheds light on the role of competition, spatial constraints, and agent traits within complex adaptive systems, offering insights into their emergent behaviors. 
\end{abstract}

\maketitle

\section{Introduction}\label{sec1}
From minute living cells to large-scale societies, complex systems are vastly diverse regarding the behavior of their constituent components and the nature of the interaction between them \cite{bibOttinoComplexSystem1,bibOttinoComplexSystem2,Meyers_2009_EncyclopediaComplex}. Driven by the interactions among the constituent components, these systems exhibit unique non-trivial properties at larger scales which is of great relevance in understanding the emergent behavior in a variety of natural phenomena \cite{bibKwapienComplexSystem, bibComplexSystemHolland}. Competition between the constituent components, aka agents, forms the hallmark of many complex systems \cite{Arthur_97_EconomyComplex, Siegenfeld2019_ComplexSystemApplications}. More specifically, in many socio-economic and biological systems, the context of interaction among the agents is set by the competition for some limited resource \cite{bibResourceAllocation, Sommer2002_CompetitionAC}. There are numerous examples of complex systems that are driven by a group of agents in common want for a shared resource \cite{bibEconophysicsChakrabarti, Arthur_94_EFBP, Cara_99_EFBP}.

It is easy to see that the active pursuit to acquire sparsely populated regions in space among a group of agents leads to competition between the agents in many complex systems \cite{bibAdaptiveMigrationSpatialCompetition}. We see the emergence of intriguing spatio-temporal patterns in social systems, such as sharing of public spaces like theaters, traffic systems, streets or squares \cite{bibSchelling1, 
 bibSchelling2}; economic systems, such as the opening of shops in an area or the mining of fuel \cite{bibSpatialEconomyGiorgio}; biological systems, such as animal foraging and in other similar spatial systems \cite{ bibFlockingAstersReversals, anand2010_ecologicalComplexity}. A number of models in which agents have spatial preferences and respond to the occupancy and behavior of agents in the neighborhood have been considered in the past. Models of cellular automata, traffic flow, swarming, flocking, opinion dynamics, spatial segregation models, etc. capture the collective behavior of a collection of agents in response to the presence and behavior of other neighboring agents \cite{TrafficFlow_Nagatani_2002, OpinionDynamics_Stauffer2009}. A particular example is the Schelling's segregation model which shows how even weak individual preferences of the agents can lead to large-scale global segregation \cite{bibSchelling1, bibSchelling2, bibSegregationNetworks}. Researchers have also explored models that examine the competitive interactions of groups of agents on spatial structures and lattices \cite{bibNowak1,bibNowak2,bibNowak3,bibHauert, bibIsingCipra, bibgamesongraph, bibshakti1}.
 
 Nevertheless, an exclusive model to study spatial spreading among agents in the quest to avoid local crowding is absent in the literature, although it is a core feature of many problems across social, economic, and biological domains \cite{Castellano2009_Stat_society, Azaele2015_Stat_Ecology, bibEconomyNetwork}. Such scenarios occur generally whenever the payoff of an agent is decided by the local spatial density of agents surrounding it and is prevalent in situations like the occupation
of shared spaces (say theaters and public transportation) or foraging activities \cite{SpaceResource_tilman1982}.  In this work, we present a simplified model that addresses the gap by focusing on a setting where space itself is the critical resource for which agents compete. The model captures essential features, such as space as a limited resource and local density-dependent payoffs. The competition among agents for the limited space then results in dynamic rearrangement in response to local crowding.

We consider a simple setting in which the aspect that space itself is the limited resource for which the agents compete is made explicit. In such competitions, often, the crowding is experienced locally. The motivation comes from the fact that,  in real space, the effect of crowding and its discomfort is predominantly felt at a local level rather than at a global scale. For instance, one might be sitting in one of the most crowded theaters, yet may feel a lesser effect of crowding if a few of her neighboring seats are vacant. Conversely, one might be in one of the least populated places and yet feel crowded if her immediate neighborhood is over-packed above her comfort level. As a result, the uncomfortable agents would seek to occupy less populated local regions in space.  This is in contrast to the typical setting considered for the competition of a scarce resource where all agents choose between the available options and the group that chose the least populated option wins \cite{bibchallet1,bibchallet2,Arthur_94_EFBP,Cara_99_EFBP}.

We use a game theoretic setting to capture the behavior of spatial systems where adaptive agents compete for a spatially spread resource. The model's rules, the pay-off structure, and agent behavior reflect the motive of the agents to avoid local crowding above their comfort level in a spatial setting.  This framework captures how the adaptive behavior of agents in pursuit of sparsely populated regions relative to their comfort level results in the emergence of interesting macro-scale behavior. In a nutshell, the model consists of agents interacting in a defined spatial environment (e.g. a lattice) with fixed rules for playing and winning.  An agent wins by occupying a neighborhood with fewer agents than her comfort threshold; otherwise, she seeks relocation to nearby regions. To isolate the core effects of competition for constrained space, we simplify the model by considering homogeneous agents and a basic spatial structure, setting aside complexities related to agent diversity and spatial heterogeneity.

The ensuing dynamics within the model unveil intriguing collective properties at a global scale. We show that the agent density significantly influences the collective behavior. Below a critical density, the system always reaches an absorbing state where all agents are winning. Above a higher density, termed the peak sustainable density, no states of the system exist wherein all agents win. We formulate an analytical method to calculate the peak sustainable density and use Monte Carlo simulations to analyze the emergent behavior of the system. We show that the spatial self-organization of the agents becomes maximally inefficient around the peak sustainable density.

We study in detail how agents' accessible information about local occupancy shapes global behavior. We show that the variation of global resource utilization with respect to the amount of information exhibits opposing trends on either sides of the peak sustainable density. We find that below the peak sustainable density, there is a non-trivial optimum value of the amount of information at which the global resource utilization is a maximum. In particular, we show that global resource utilization can be a non-monotonic function of the amount of information.  Intriguingly, possessing more information proves counterproductive for better global outcomes at these densities. We also probe how the inequality between the agents measured in terms of the difference in their total payoff obtained over the course of the game is affected by the density and information. Unlike inefficiency, we find that inequality, in general, shows a decreasing trend with increasing information. Thus, analyzing the macro-scale properties, we showcase how agents' crowd-avoidance and relocation lead to novel emergent behavior. Our results uncover scenarios where local crowd-avoidance prompts global crowding, and increased individual-level information yields diminished global efficiency - defying conventional expectations.

\par The paper is organized as follows. In Section~\ref{Model}, we introduce the model and define the terminologies and relevant macroscopic parameters of the system. We also present calculation of the peak sustainable density in the system. In Section \ref{Analysis}, we present the simulation results regarding the macroscopic properties like the global inefficiency, inequality, and fraction of permanent winners in the system. We discuss in detail their dependence on the parameters of the model. Section \ref{Conclusion} contains the conclusion and discussion.

\section{The model} \label{Model}

In the model, agents engage in competition to secure sparsely populated neighborhoods, guided by their individual comfort thresholds. We consider the space as a lattice upon which the agents are distributed. For conceptual simplicity, we consider the one-dimensional (1D) lattice consisting of  \textit{L} sites where $N$ agents are randomly distributed initially ($N \leq L$). Each site can be occupied by only one agent at a time. Periodic boundary conditions are assumed to reduce edge effects. System's \textit{Density} ($\rho$) is defined as the ratio of the total number of agents ($N$) to the number of sites ($L$). In the following, we introduce the main parameters of the model and discuss the dynamics. 

The \textit{neighborhood of a lattice site} is constituted by the physically nearest sites of it. In one dimension, the simplest neighborhood consists of the two immediate neighboring sites to the left and right of a focal location. The \textit{size of the neighborhood ($z$)} in this case is 2. Advancing this notion,  $z=4$ corresponds to the case where the neighborhood comprises the four nearest sites, two on each side of a site (for simplicity, we will consider only neighborhoods that are symmetric to the right and left of a site). The success of an agent is assumed to depend on the \textit{occupation density ($\nu$)} within her neighborhood. For the $i$th site, the occupation density of its neighborhood ($\nu_i$) is defined as the ratio of the number of agents occupying the neighborhood to the number of sites in the neighborhood.

\par The \textit{tolerance threshold} ($\tau$) is the limit of the occupation density within the neighborhood of an agent's location, up to which she is a winner (payoff = 1). When the occupation density crosses the tolerance threshold, an agent will turn a loser (payoff = 0) and seek options for relocation. For example, for a 1D lattice with $z=2$, $\tau$ can adopt values of 0 (indicating agents' intolerance to any neighborhood occupancy), $0.5$ (indicating tolerance for up to $50\%$ neighborhood occupancy), or 1 (indicating tolerance for $100\%$ neighborhood occupancy). Clearly, for $\tau=1$, all agents are winning all the time. While, in general, $\tau$ can vary among agents or even across sites, we focus on the simplest scenario where $\tau$ remains uniform across all agents and sites. 

The notion of neighborhood size and tolerance threshold encapsulates the concept of competition for space in the model. Together, they determine the level of tolerance of agents towards local crowding. The essence of the crowd avoidance rule is contained in the fact that when an agent chooses a region in space whose occupation density is less than or equal to the tolerance threshold, she is winning.

\begin{figure}
\includegraphics[width= 0.6\textwidth]{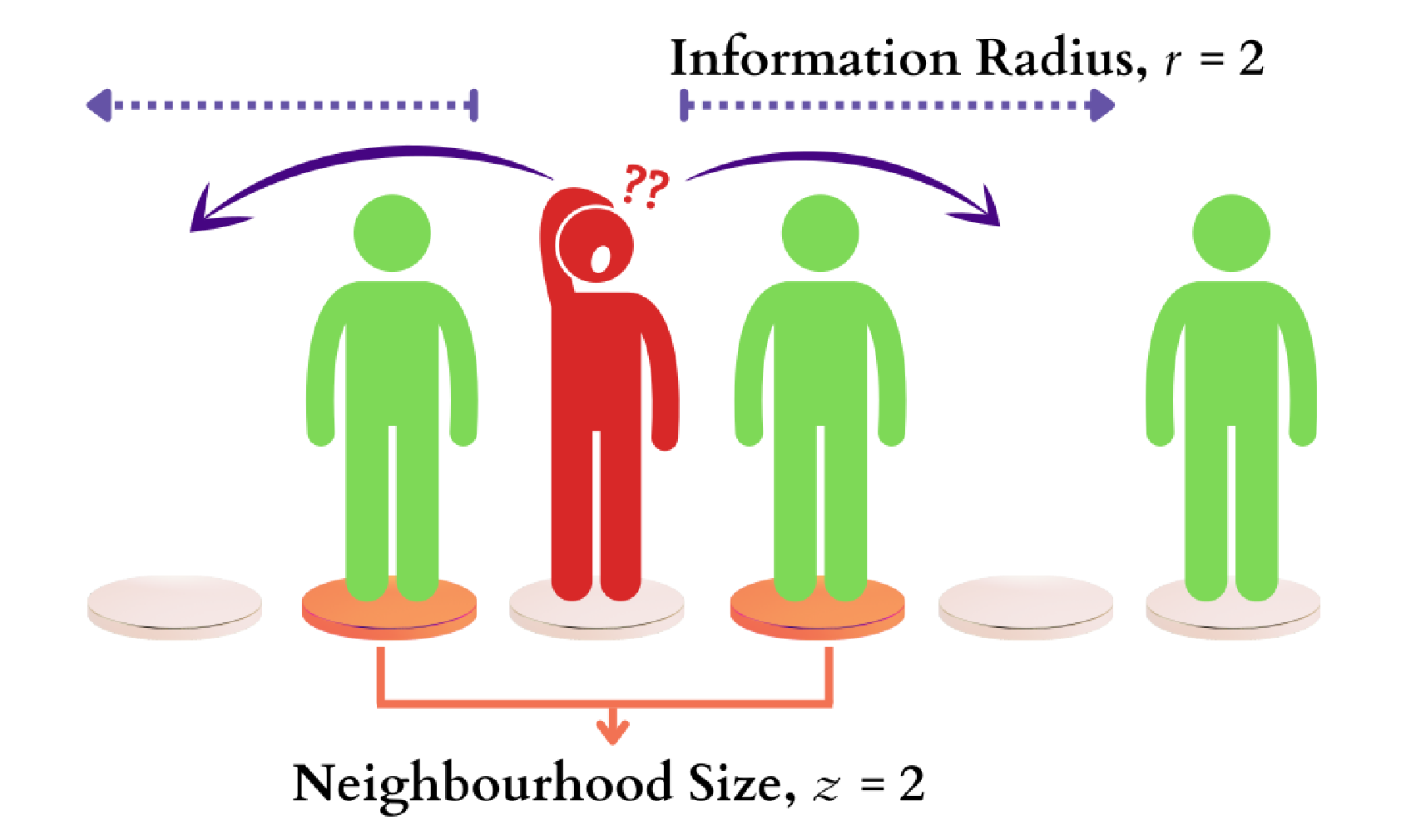}
\caption{\label{fig:Game} A schematic illustration of the decision-making process of an agent in the model for a $z=2; \tau= \frac{1}{2}; r = 2$ system showing winning (green or lighter color) and losing (red or darker color) agents for a particular round of the game. Agents are placed on a 1D lattice with only singe occupancy possible at a site. Winners stay put, and losers try to relocate to a vacant site within a radius of $r$. The available relocation options for the losing agent are denoted by curved arrows.}
\end{figure}

In the model, an agent is assumed to have access to information about the occupancy of all lattice points up to \textquoteleft$r$\textquoteright\; sites on either side from her present location, defined as the \textit{Information radius}. Alternatively, we consider the standardized information radius ($r_s$), which is the ratio of the information radius ($r$) to $z/2$, thus normalizing the information radius with respect to the neighborhood size. Again, for simplicity, the value of $r$ is assumed to be the same for all agents. An agent uses information about the occupancy within her information radius to decide what to do in the subsequent round of the game.  Consequently, an agent's movement is assumed to be confined within this radius, given the absence of information beyond this limit. The case where the information radius is smaller than the neighborhood radius (i.e., $r<z/2$ or $r_s < 1$) corresponds to situations where the range of movement of an agent is limited even though the agent’s level of comfort (payoff) is determined with respect to a larger neighborhood. In real systems, such limitations may occur due to factors like cost of movement or relocation to faraway points. This cost could be due to the effort, time, and energy required to relocate/move or due to the resources required to gather information about the faraway vacant sites.

\begin{figure}
\includegraphics[width=0.48\textwidth]{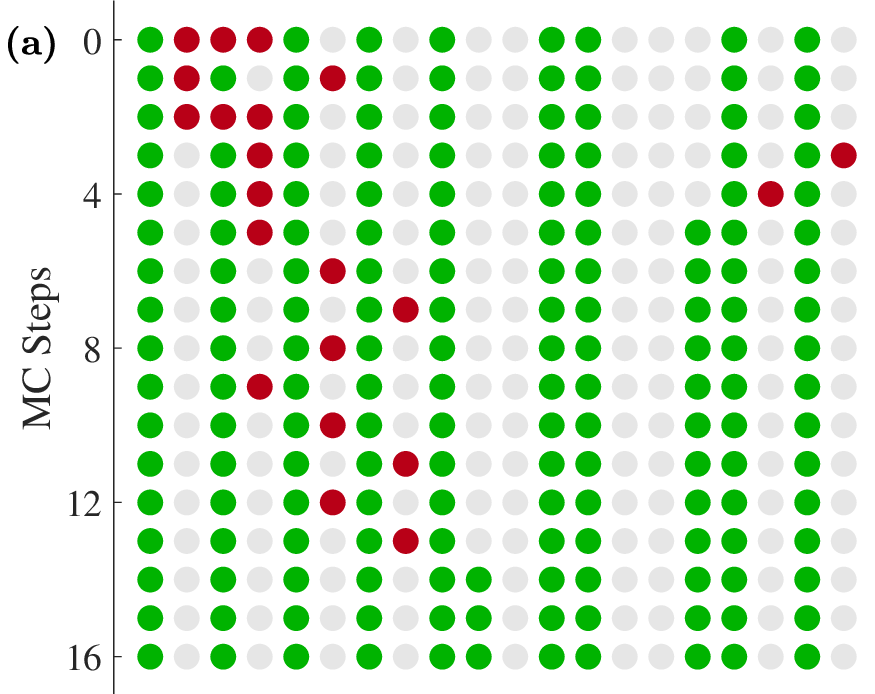}\hfill
\includegraphics[width=0.48\textwidth]{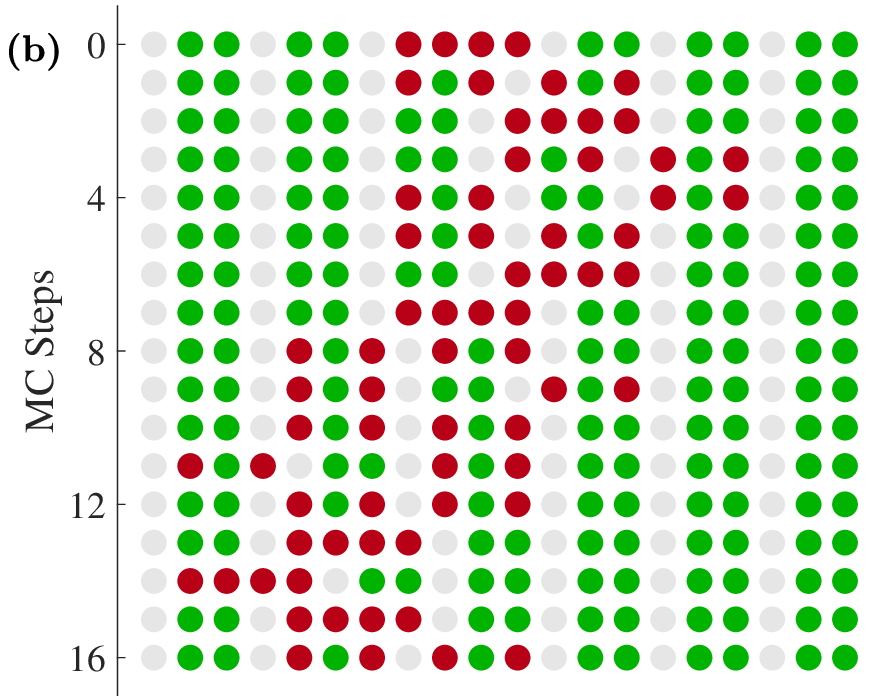}
 \caption{\label{fig:Agent_Movement_and_Dynamic_Interactions}Dynamic progression of agent configurations in the 1D model. Each row illustrates the positions of winning agents (green), losing agents (red), and vacant sites (light grey) at a given time step. \textbf{(a)} depicts the time evolution of a system with $z=2$, $\tau=\frac{1}{2}$, and $r=2$, showing the first 16 Monte Carlo (MC) steps from a random initial configuration. An absorbing state is reached by the 14th step. \textbf{(b)} shows the time evolution of a system with $z=4$, $\tau=0.5$, and $r=3$, after $10^{5}$ initial steps, where the dynamics remain unsettled.}%
\end{figure}

The adaptive behavior of the agents is ingrained in the mechanisms they adopt at the individual level to respond to the local environment. 
All agents follow a simple win-stay, lose-shift behavioral rule \cite{bib_Win_Stay_Lose_Shift}. The winners do not move. i.e., they remain at their current locations for the next round of the game. Each loser attempts to relocate to any of the vacant sites within her information radius with equal probability. If no vacant sites exist within this radius, the losing agent remains stationary. We employ a sequential update scheme in which, at each step, an agent is chosen randomly to update its location. Fig.~\ref{fig:Game} gives a schematic illustration of the decision-making process of an agent, and Fig.~\ref{fig:Agent_Movement_and_Dynamic_Interactions} depicts typical examples of the game's dynamic progression across sequential Monte Carlo steps, where each row represents the arrangement of agents at one instance.  A Monte Carlo step (MC step) consists of one move made by a randomly selected losing agent. The states in which all agents are winners are the absorbing states of the system since no agents would move any further. Thus, the absorbing states can only be reached but not left, breaking the condition of detailed balance. The existence of absorbing states makes the overall dynamics a non-equilibrium process \cite{Hinrichsen_nonequilibrium_1,Hinrichsen_nonequilibrium_2}.

The agents carry an accumulated payoff that contains their long-term winning history. For each agent, the payoff structure adds a score of one for a winning round and a score of zero for a losing round. Therefore, the cumulative payoff of an agent who is at the $i$th site after $t$ rounds can be written as 

\begin{equation}
P(t) = 
\begin{cases}
    P(t-1)+1 & \text{ if } \nu_i(t)\le \tau\\
   
    P(t-1) & \text{ if } \nu_i(t) > \tau      
\end{cases}
\label{eqn: payoff}
\end{equation}

The cumulative payoff thus measures an individual's long-term wealth and success rate. All the agents start with a score of zero. The motive of each agent is to maximize their personal payoff through relocation and is assumed to be indifferent to the welfare of neither the other agents nor the system. 

The model can be generalized to higher dimensions by considering higher-dimensional lattice structures and neighborhoods. For example, in two dimensions, which may be of particular relevance for many real-world systems, we may consider a square lattice, with the neighborhood following a Von Neumann structure. For a neighborhood radius of 1, the neighborhood includes the four adjacent sites: up, down, left, and right. As the radius increases, it expands to include sites reachable within fewer than $\frac{z}{2}$ steps, with movement at each step restricted to the four directions. Similarly, the information radius includes all sites reachable within $r$ steps.

We also note that although our current model is defined on a one-dimensional lattice with corresponding neighborhoods and information radii, the agents can jump over occupied lattice sites which somewhat dilutes the one-dimensional nature of it. As a result, we expect the behavior of the two-dimensional model to be somewhat similar to the one-dimensional case. For smaller neighborhood sizes, we confirmed that the best packing density (see Sec.~\ref{EmergentProperties}) is identical in both one and two dimensions, and the behavior of the emergent macroscopic properties is largely comparable. A detailed analysis of the two-dimensional model is not attempted here.

\subsection{Emergent macroscopic properties \label{EmergentProperties}}

The local crowd avoidance in the model leads to collective behavior at the global level. In this respect, an interesting question is whether the agents will achieve emergent coordination leading to optimal occupation of sites. Another aspect is how the winning chances are distributed in the population. To characterize the large-scale behavior of the system, we consider the emergent properties of global inefficiency, inequality, and fraction of frozen agents which are defined below.

 \textit{Global inefficiency ($\eta$)} is defined as the normalized deviation of the number of losers from its minimum possible value averaged over several rounds in the steady state and for several initial conditions.  If $N_l$ is the random variable representing the number of losers found in the system for a round of the game and $N_{l_{min}}$ is its minimum possible value, we can define the global inefficiency per agent by, 
 \begin{equation} 
\eta = \frac{\langle\langle N_l - N_{l_{min}} \rangle\rangle}{N}
\label{eqn: inefficiency}
 \end{equation}
 
Here, the double angular bracket indicates averaging over several rounds in the steady state and also over several initial conditions. $N_l = N_{l_{min}}$ corresponds to the best possible arrangement of agents in the system for a given set of parameters. It is easy to see that $N_{l_{min}}$ will be zero for very low densities and it will approach $N$ as the density approaches one. Inefficiency $\eta$ thus quantifies the degree of emergent organization in the system. In other words, it is a measure of the wastage of resources in the system at a global scale.

In the model, inequality between the cumulative payoffs of different agents may arise. Losers may remain losers repeatedly, and winners remain winners. We use the \textit{Gini coefficient} to measure the level of inequality in the system \cite{bibGini1}. If $P_k$ denotes the wealth of the $k^{th}$ agent and $\Bar{P}$ denotes the mean wealth of all the $N$ agents in a round, then the Gini coefficient for a round is given by

\begin{equation} 
    g=\frac{\displaystyle\sum_{k=1}^{N} \displaystyle\sum_{l=1}^{N} \mid P_k-P_l\mid}{2\displaystyle\sum_{k=1}^{N}\displaystyle\sum_{l=1}^{N}P_k}
    =\frac{\displaystyle\sum_{k=1}^{N} \displaystyle\sum_{l=1}^{N} \mid P_k-P_l\mid }{2N^2\Bar{P}} 
\label{eqn: Gini Coefficient}
\end{equation} 

As in the case of $\eta$, the Gini coefficient $G$ for a particular set of parameters is then found by averaging $g$ over several rounds in the steady state and several initial conditions. The value of $G$ ranges from 0 to 1, with 0 representing perfect equality and 1 indicating perfect inequality.

A different quantity to characterize the inequality is the fraction of frozen agents in the system ($\phi_w$). i.e., agents who remain permanent winners and therefore remain stationary throughout the duration of the game. As we'll see, $\phi_w$ also characterises the order-disorder transition in the system.

\subsection{Peak sustainable density} \label{Peak_Sust_Density}
We first address the following question: What is the maximum number of agents that can be packed or the maximum density that can be achieved in a 1D lattice such that the occupation densities inside the neighborhood of all the agents are less than their tolerance threshold (i.e., all agents are winning)? The answer to this can give us useful insights into the dynamics of the system for given values of $z$ and $\tau$. More importantly, this will give us the value of $N_{l_{min}}$, which is required to calculate global inefficiency (Eq.~\ref{eqn: inefficiency}).

 We define \textit{peak sustainable density}, $\rho_p$, as the maximum density of agents possible in the system with all the agents in a winning state and would like to determine it as a function of $z$ and $\tau$. Now, the densest packing of agents is attained when a maximum number of agents share each vacant site, and each agent accommodates the maximum possible number of agents in her neighborhood, still being a winner. To keep things manageable, we restrict our search to configurations with clusters of agents assuming equal size. A cluster of agents is defined as a maximal set of contiguous agents with vacant sites surrounding it. An exhaustive search for relatively small system sizes and neighborhood sizes confirms that in most cases, the densest packing is obtained for configurations with uniform-sized clusters of agents. In the following, we describe a systematic procedure for finding this dense packing, assuming that the clusters are of equal size. 
 
\begin{figure*}
\centering
\includegraphics[width= 0.8\textwidth]{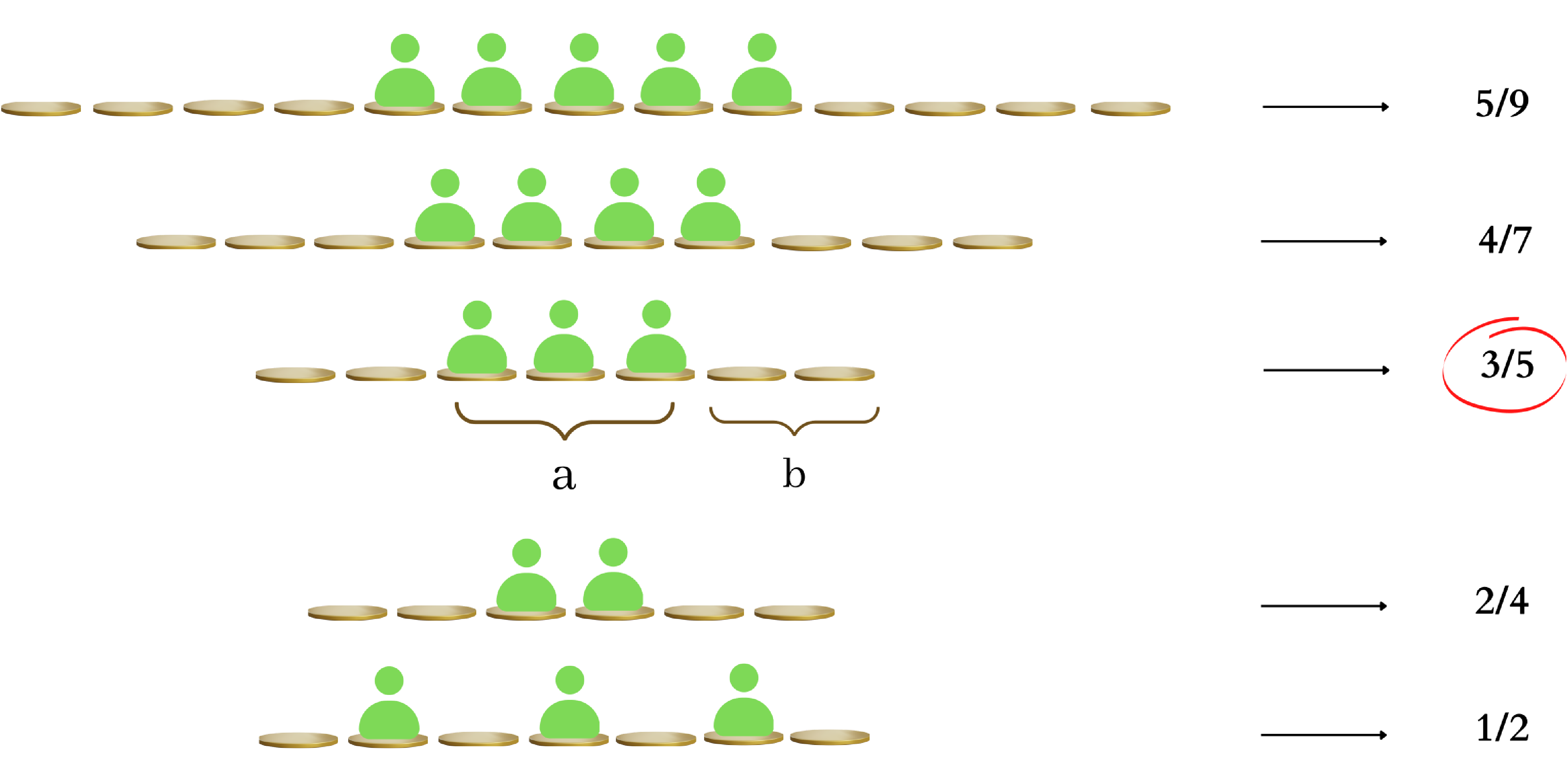}
\captionsetup{justification=justified,singlelinecheck=false, width=\textwidth}
\caption{A schematic illustration of the different possible configurations for a $z=8$, $\tau=\frac{4}{8}$ system for different cluster sizes such that all the agents in the clusters are winners. Vacancies are added, ensuring that each agent in the cluster has at least $z-\tau z=4$ vacant sites within her neighborhood. To create a system of size $L$, clusters and cluster gaps can be added consecutively (assuming that $L$ is a multiple of (Cluster Size $+$ Gap Size)). This process results in configurations with clusters of equal size and spacing, with all agents winning as shown in different rows. The density of the resulting configurations shown in the figure towards right are calculated using Eq.~\ref{eqn:Peak_Sust_density}. The cluster size and cluster gap size of the densest packing are denoted by $a$ and $b$, respectively, and its density gives the peak sustainable density, $\rho_p = 3/5$ (encircled in the figure).}
\label{fig: Clustering_Replication}
\end{figure*}

First, form clusters of agents on the lattice with a minimal but sufficient number of vacancies in between the clusters such that all the agents are winning. Within the neighborhood size $z$, each winning agent can afford a maximum of $\tau z$ number of agents. Hence, the biggest possible cluster of winning agents has a size of $\tau z +1$. Obviously, the smallest possible cluster size is one. Now consider all possible configurations with cluster sizes varying from the maximum to the minimum possible value, dispersing vacancies (cluster gaps) in between, ensuring that all the agents have sufficient vacancies within their neighborhood to be winners. Among the different configurations that can be generated in this way, we can choose the one with the highest value of density ($\rho$).

 To illustrate the above, consider, for example, a system with a neighborhood size, $z=8$ and tolerance threshold, $\tau=\frac{4}{8}$. Each agent requires at least $z- \tau z=4$ vacant sites in her neighborhood to be a winner. Fig.~\ref{fig: Clustering_Replication} shows the various possibilities of forming clusters of winning agents with minimal but sufficient vacancies around the cluster. Clusters vary in size from the maximum possible value of five to the minimum possible value of one. The densest configuration is found to be the one with a cluster size of three and a cluster gap size of two. 

As the cluster size is incrementally reduced from the largest possible value, the neighborhood of the agents starts to extend into the nearby clusters and cluster gaps. If removing one agent from the cluster results in all the other agents gaining an additional vacant site in their neighborhood, then the cluster gap size too can be reduced accordingly. For example, in Fig.~\ref{fig: Clustering_Replication}, when the cluster size is reduced from five to four, the cluster gap size can be reduced by one. On the other hand, if the removal of one agent results in the neighborhood of any one of the agents in the cluster extending into an occupied site in a nearby cluster, the cluster gap size can not be reduced. Notice that the cluster gap size remains stagnant at two when the cluster size is reduced from three to two in Fig.~\ref{fig: Clustering_Replication}. 

Peak sustainable density is the highest value of density for the configurations generated in this manner. If the cluster size and cluster gap size for the densest configuration are denoted by $a$ and $b$, respectively, then the peak sustainable density is given by

\begin{equation}
\rho_p = \frac{a}{a+b}
\label{eqn:Peak_Sust_density}
\end{equation}

\subsubsection{General expression for peak sustainable density}

We formulate a general expression for the peak sustainable density in terms of $z$ and $\tau$, following the method described above. Initially, when the biggest possible cluster of size $\tau z +1$ is formed, the requirement of having $z-\tau z$ vacant sites in the neighborhood of each winning agent necessitates placing cluster gaps of size $z-\tau z$ on both sides of each cluster. Even so, the cluster gap size need not be greater than $\frac{z}{2}$, since the neighborhood of the agents extends only up to this limit towards either side of an agent. As a result, we have two scenarios that should be treated separately. 

\begin{enumerate}

\item For $z-\tau z > \frac{z}{2}$, (i.e., $\tau < \frac{1}{2}$), the minimum cluster gap size needed for the biggest possible cluster with all agents winning is fixed at $\frac{z}{2}$. In such configurations, all the agents in a cluster can access the two neighboring cluster gaps on either side, even for the biggest possible cluster. 

\item For $z-\tau z \leq \frac{z}{2}$, (i.e., $\tau \geq \frac{1}{2}$), the minimum cluster gap size required for all the agents in the biggest possible cluster to be winning  is $z- \tau z$, a function of $\tau$. In this case, a few agents of the biggest possible cluster can access only one of the neighboring cluster gaps. 

\end{enumerate}

For $\tau < \frac{1}{2}$, we find that the densest possible packing with all agents winning is when the agents are packed as clusters of the largest possible size with a cluster gap of $\frac{z}{2}$ in between (see Appendix~\ref{AppendixA} for details). Hence, the peak sustainable density for $\tau < \frac{1}{2}$ is given by

\begin{equation}
\rho_p= 
    \frac{\tau z+1}{\tau z+1+ z/2} 
    \label{eqn:PeakDensity_tau_lessThanHalf}
\end{equation}

For $\tau \geq \frac{1}{2}$, we see that there are two candidate arrangements of agents that will result in the highest density with all agents winning (See Appendix~\ref{AppendixA} for details). The first possibility is when the agents are arranged as the biggest possible clusters with a cluster gap of size $ z-\tau z$ in between. In this scenario, the neighborhoods of some agents within a cluster extend only to a single cluster gap. The packing density in this case is given by

\begin{equation}
\rho =  \frac{\tau z+1}{z+1} 
\label{eqn:PeakDensity_tau_greaterThanHalf_a}
\end{equation}

The second possibility involves forming the largest possible cluster where the neighborhoods of agents at the cluster's edge span entirely across the two nearest cluster gaps, one on each side. For such a scenario, the packing density is given by

\begin{equation}
\rho = 
    \frac{\frac{z}{2}-j_{max}}{(\frac{z}{2}-j_{max})+\lceil \frac{z-\tau z}{2} \rceil } 
\label{eqn:PeakDensity_tau_greaterThanHalf_b}
\end{equation},

where $j_{max}$ is

\begin{equation*}
j_{max} = \lfloor \frac{z-\tau z}{2}  \rfloor-1 
\end{equation*}

Here, $\lceil x \rceil$ and $\lfloor x \rfloor$ represent the ceil and floor functions, which, respectively, round a real number $x$ up and down to the nearest integer. The peak sustainable density for $\tau \geq \frac{1}{2}$ is the highest value among the two fractions given by Eqs.~\eqref{eqn:PeakDensity_tau_greaterThanHalf_a} and \eqref{eqn:PeakDensity_tau_greaterThanHalf_b}. Peak sustainable density values for a few values of $z$ and $\tau$ determined using the method described above and verified by exhaustive search are listed in Table~\ref{table:Critical_Density}.

\begin{table*}
\centering
\renewcommand{\arraystretch}{1.3}
  
\begin{tabular}{l | l l l l | r r r r } 
\hline
$z$ & $\tau =0$ & $\tau =\frac{1}{z}$ & $\tau =\frac{2}{z}$ &$\tau =\frac{3}{z}$ & $\tau =\frac{z-4}{z}$ & $\tau =\frac{z-3}{z}$ & $\tau =\frac{z-2}{z}$  & $\tau =\frac{z-1}{z}$  \\ [0.5ex] 
\hline
\hline
 2 & 1/2 & -   & -    & -    & -    &-    & -  & 2/3   \\ 
 4 & 1/3 & 2/4 & -   & -     & -    &-     & 2/3 & 4/5  \\
 6 & 1/4 & 2/5 & 3/6 & -     & -    & 3/5  & 3/4  & 6/7 \\
 8 & 1/5 & 2/6 & 3/7 & 4/8   & 3/5 & 4/6 & 4/5 & 8/9 \\
  [1ex]

\hline
\end{tabular}
\caption{Peak sustainable densities for a few values of neighborhood sizes $z$ and tolerance threshold $\tau$.}
\label{table:Critical_Density}
\end{table*}

 \subsubsection{Guaranteed number of losers}
 
\begin{figure*}
    \centering
    \includegraphics[width= 0.8 \textwidth]{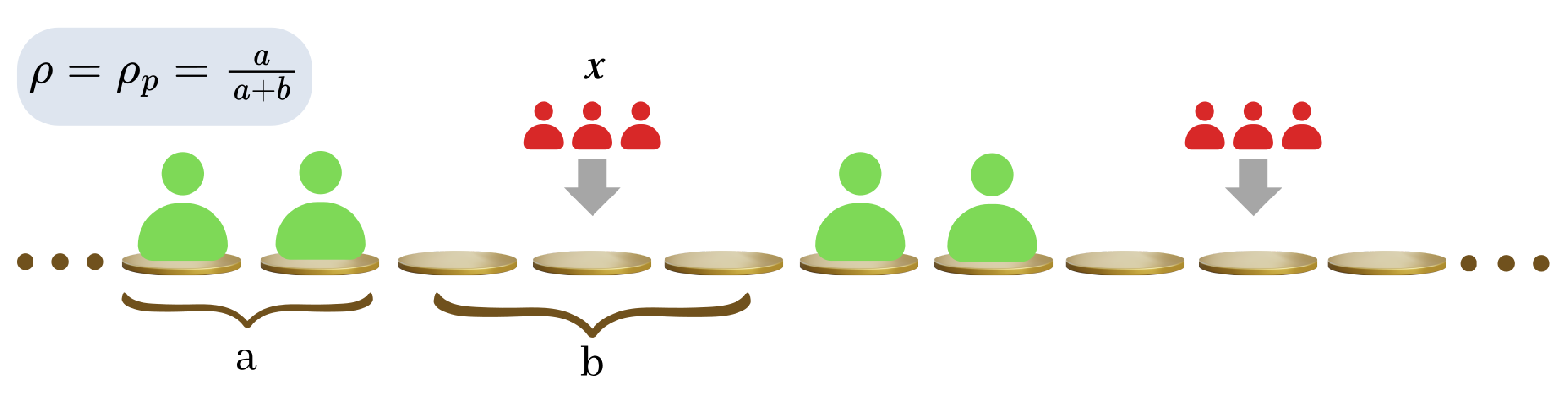}
    \caption{The arrangement of agents corresponding to the peak sustainable density $(a=2; b=3; \rho_p= \frac{2}{5})$ for $z=6; \tau= \frac{1}{6}$. Adding extra agents ($x$) above the peak sustainable density (indicated by the dark red agents) will create losers in the system as given by Eq. \ref{MinLosersEqn}.}  
    \label{fig:Extra_Agents}
\end{figure*}

In the model, the states of the system in which all the agents are winners and hence cease to have further dynamics are absorbing states. For low values of density, such absorbing states are aplenty. However, as the density increases, the configurations where everyone is a winner become fewer in number. Above the peak sustainable density ($\rho_p$), there are no such absorbing states, and a non-zero number of losers will always be present in the system. Thus, the peak sustainable density signals the change in the value of the minimum possible number of losers ($N_{l_{min}}$) from zero to non-zero.

Let $x$ denote the number of extra agents present in the system above the number of agents corresponding to the peak sustainable density ($\rho_p \times L $). i.e.,

\begin{equation*}
 x = 
\begin{cases}
    0& \text{if } \rho \le \rho_p\\
   
    (\rho-\rho_p)L& \text{if } \rho > \rho_p      
\end{cases}
\label{eqn: x}
\end{equation*}

When such extra agents are present, the configurations with the maximum number of agents winning are achieved when the population is arranged in clusters and cluster gaps of size $a$ and $b$, respectively, and the extra agents are filled in nearby cluster gaps. An example is shown in Fig.~\ref{fig:Extra_Agents}. Note that adding extra agents in this way will render some existing agents as well as the newly added ones losers. We can say that, in general, adding $b$ number of extra agents into the system creates approximately $(a+b)$ number of losers. Therefore, the expression for the minimum guaranteed number of losers in the system can be written as 

\begin{equation}
     N_{l_{min}} \approx (a+b) \frac{x}{b}
     \label{MinLosersEqn}
 \end{equation} 
We use this value to calculate the global inefficiency in Eq.~\eqref{eqn: inefficiency}.

\section{Simulation results} \label{Analysis}

\begin{figure}
\includegraphics[width=0.6\textwidth]{ 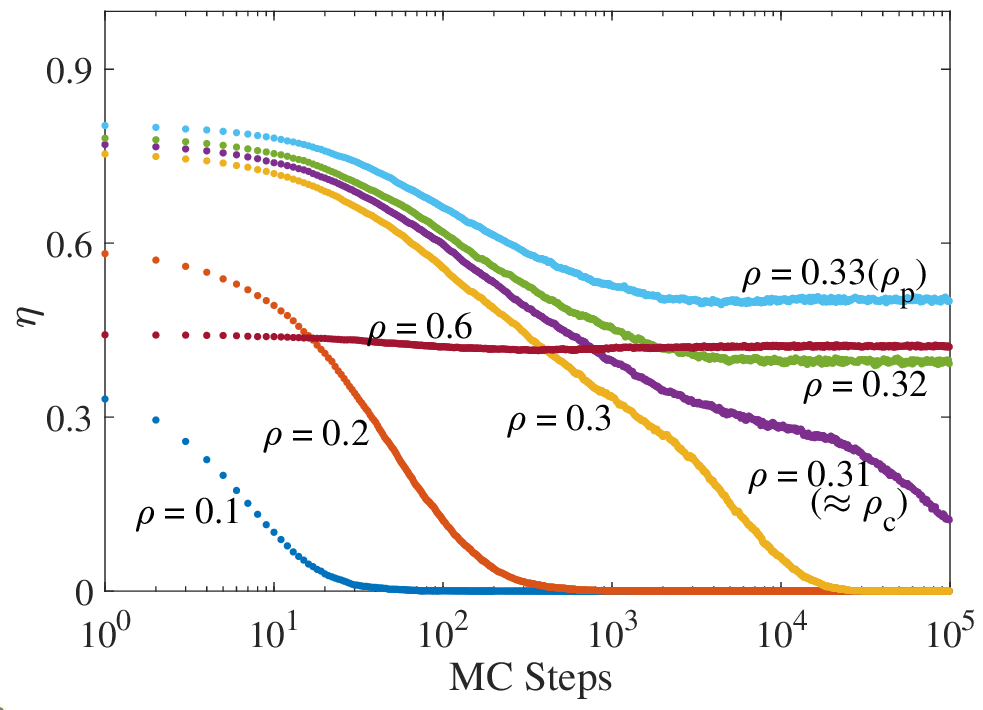}
\caption{Evolution of inefficiency per agent ($\eta$) with respect to Monte~Carlo steps for an $L=201; z=4; \tau =0$ system with $r=3$.  For a very narrow density window near a critical value, the relaxation time diverges to infinity ($ \rho \approx 0.31$ corresponds to this critical value of density in this case). For densities above and below this narrow range, the system reaches its steady state faster (typically in less than $10^4$ MC steps for the values of the parameters used).}
\label{Time_steps}
\end{figure}

We simulate the model using Monte Carlo methods and investigate its emergent behavior. For a given system with specific values of $z$ and $\tau$, two factors that play a major role in determining its dynamics and evolutionary behavior are density ($\rho$) and information radius ($r$). We study the steady-state statistical properties of the system, focusing on the relation of the emergent properties to these two parameters. Specifically, we focus on the macroscopic properties of the system, such as global inefficiency, the fraction of frozen agents, and the inequality measured by the Gini coefficient. 

The simulations start with a random distribution of agents on the lattice, and the results discussed in the subsequent sections use a system size of $L=201$, unless otherwise specified. We chose $L=201$, an odd integer, to avoid the repetition of neighboring sites with periodic boundary conditions when the neighbourhood span the system size, $\frac{z}{2}=\lfloor \frac{L}{2} \rfloor$. The primary influence of system size on macro-observables depends on whether $L$ is an exact multiple of the optimal cluster size plus cluster gap size $(a+b)$. Depending on the remainder of $\frac{L}{(a+b)}$, the optimal number of agents that can be packed onto the lattice may vary slightly. However, this effect becomes negligible when L is significantly larger than (a+b). Thus, the qualitative features of the system measured show negligible dependence on the system size $L$ for sufficiently large $L$. As visible in Fig.~\ref{Time_steps}, for lower densities, the inefficiency reaches the steady state value of zero in a few MC steps. At higher densities, the system stabilizes to the steady state value in approximately around $10^4$ steps. Relaxation time diverges as we approach a critical density which is defined later.
We measure the macroscopic properties of the system over the final $10\%$ of the total $10^4$ Monte Carlo steps. The properties were then averaged over $10^3$ random initial distribution of agents on the lattice.

\subsection{Population density and inefficiency} \label{Density_and_Inefficiency}

\begin{figure*}
\includegraphics[width=.49\linewidth]{ 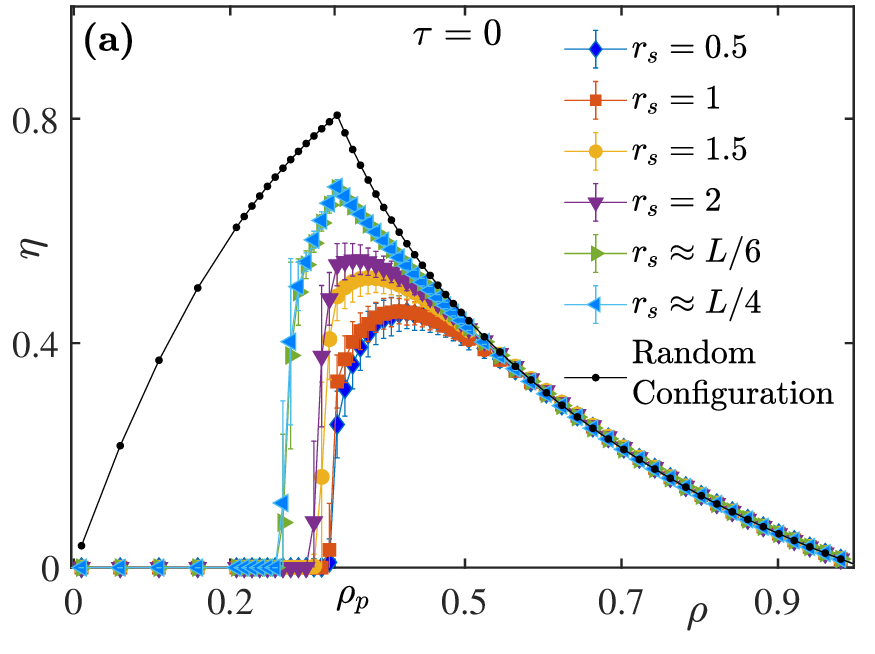}
\includegraphics[width=.49\linewidth]{ 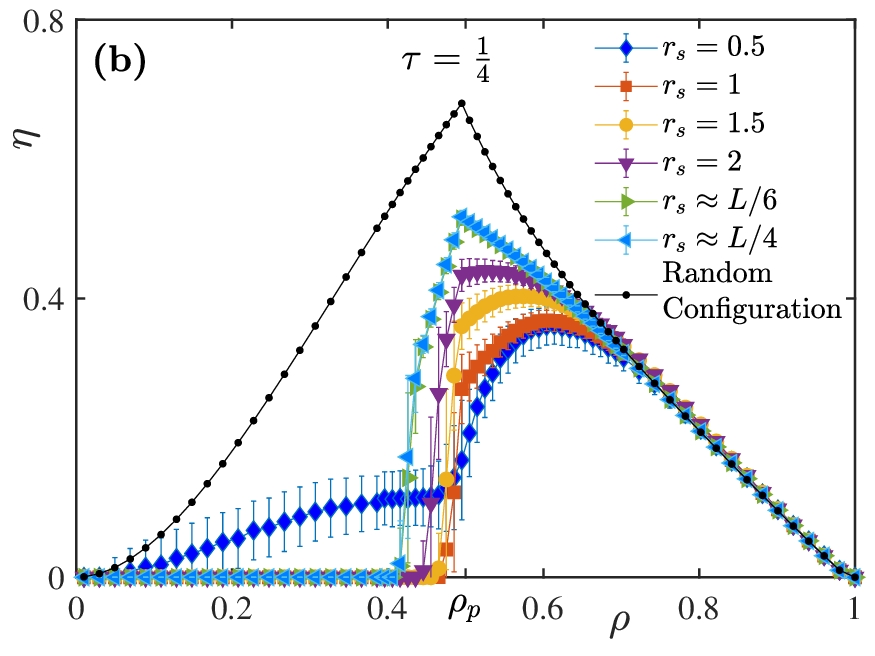}
\includegraphics[width=.49\linewidth]{ 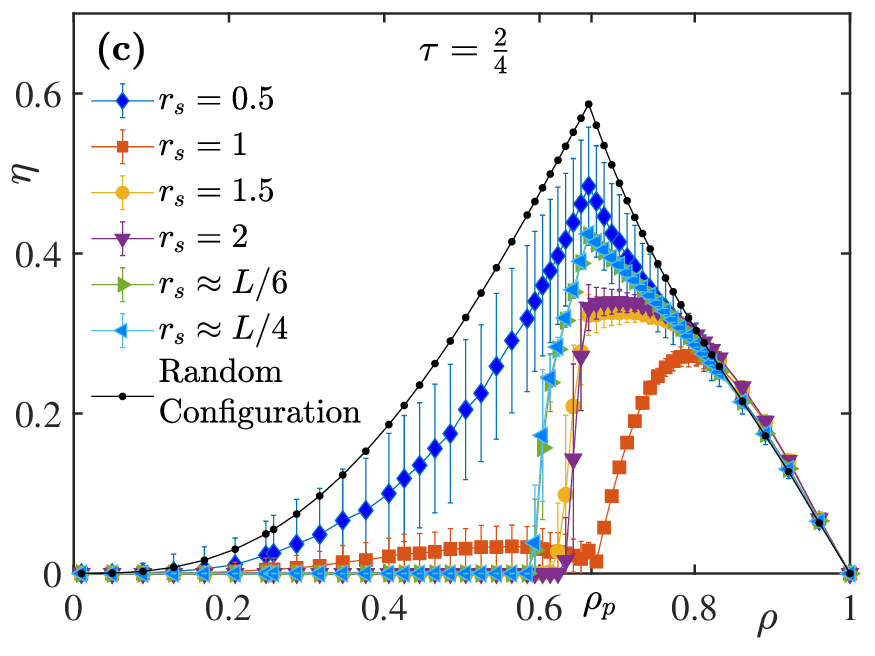}%
\includegraphics[width=.49\linewidth]{ 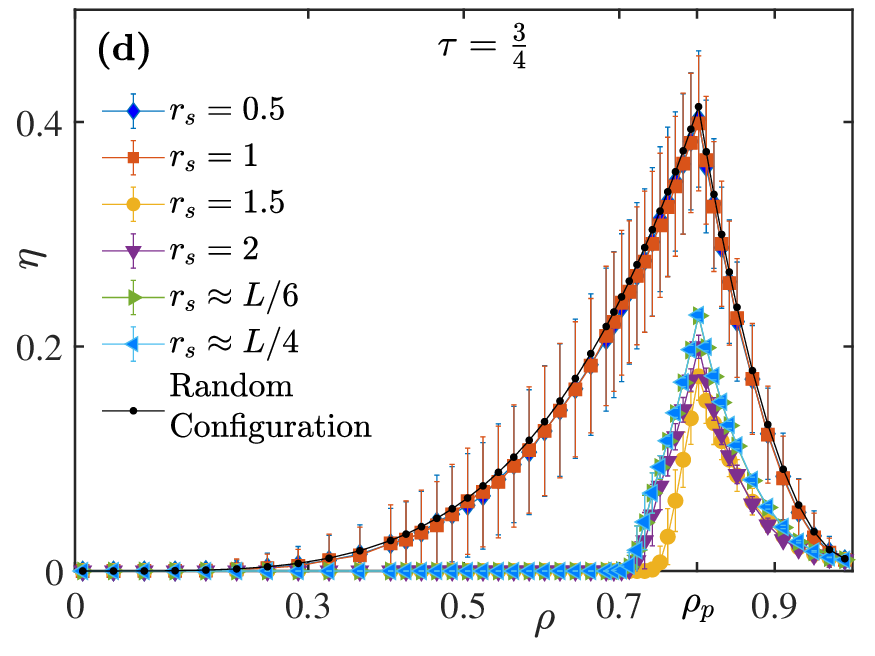}%
\caption{Variation of inefficiency ($\eta)$ with density ($\rho$) for $z=4$ system for different standardized information radii ($r_s$) and tolerance thresholds  \textbf{(a)} $\tau=0$, \textbf{(b)} $\tau=0.25$, \textbf{(c)} $\tau=0.5$ and \textbf{(d)} $\tau=0.75$. The black curve with dots corresponds to the inefficiency of a random distribution of agents on the lattice. In all cases, for  $r_s > min\{ 2 \tau, 1\}$, inefficiency turns non-zero at the critical density ($\rho_c$) and peaks near the peak sustainable density ($\rho_p$). System size used is $L=101$.}
\label{fig:Inefficiency_Vs_Density}
\end{figure*}

The population density is an important determinant of the evolution of the system and impacts the inefficiency and other macroscopic properties. Now, the inefficiency of a random arrangement of agents can be easily found to be
\begin{equation} \label{eqn: Inefficiency Random configuration}
 \eta_{\text{rand}} = \displaystyle\sum_{i=0}^{z-\tau z-1} \binom{z}{i}\rho^{(z-i)}(1-\rho)^i-\frac{N_{l_{min}}}{N}       
\end{equation}
where the summation is over the number of vacant sites in the neighborhood of an agent and $(z-\tau z-1)$ is its maximum value such that the agent is a loser. The adaptive behavior of the agents facilitates the reduction of inefficiency from its value of a random configuration. The variation of inefficiency with density for different values of the standardized information radius $r_s = 2r/z$ and different $\tau$ values for a $z=4$ system are shown in Fig.~\ref{fig:Inefficiency_Vs_Density}. 

We define critical density ($\rho_c$) as the density below which a system can self-organize and achieve zero inefficiency. Below $\rho_c$, the system always reaches one of its absorbing states. $\rho_c$ can be considered as the \textquoteleft carrying capacity\textquoteright\; of the system demarcating the transition from an efficient to an inefficient phase. The value of critical density depends on system parameters of neighborhood size, tolerance threshold, and information radius. 

\begin{figure}
\includegraphics[width=0.48\textwidth]{ 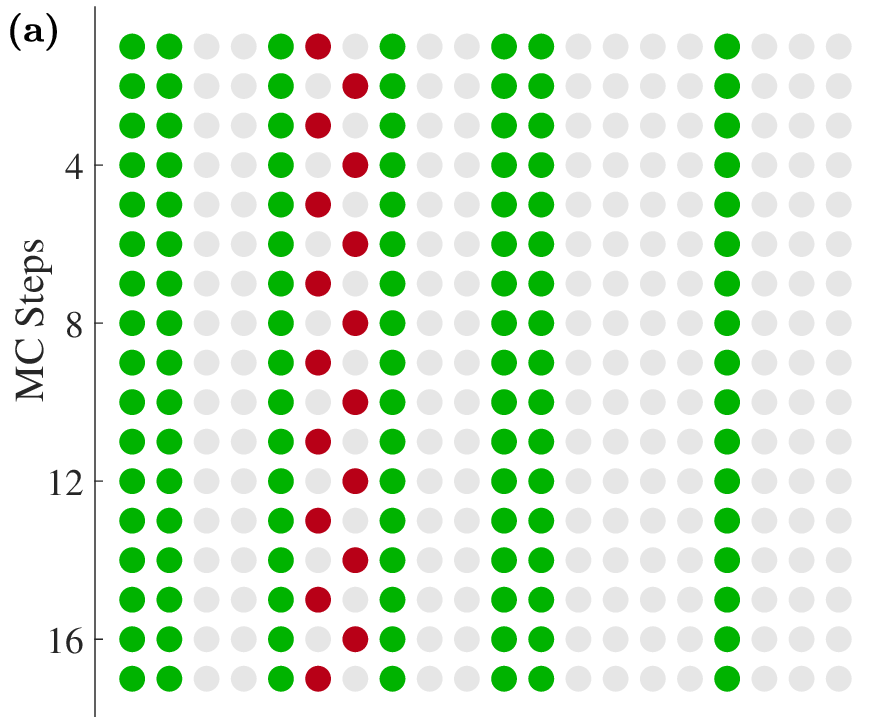}
\includegraphics[width=0.48\textwidth]{ 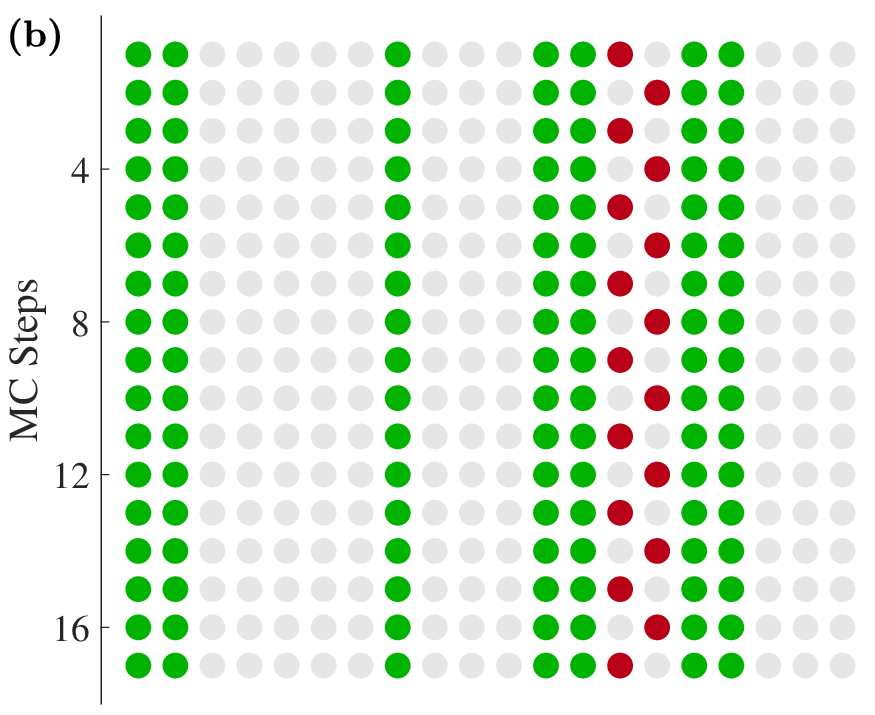}
\caption{\label{fig:Trapping_States}%
Visual representation of typical trapping states in the dynamics for \textbf{(a)} $z=4; \tau = \frac{1}{4}$ system for $r_s=0.5$ and \textbf{(b)} $z=4; \tau = \frac{2}{4}$ system for $r_s=1$. Light grey, green and red dots denote vacant sites, winning and losing agents, respectively. The trapping states arise when the losing agents are trapped between stationary winning agents, which may happen when $r_s \leq min\{2 \tau, 1\}$.}  
\end{figure}

From Fig.~\ref{fig:Inefficiency_Vs_Density}, it is clear that for certain systems, the value of critical density is non-zero and closer to $\rho_p$, and for certain other systems, its value tends towards zero. This is because, for a given $\tau$, whenever $r_s \leq min\{ 2 \tau, 1\}$, a losing agent between two clusters of winning agents on either side, each of size $r = \tau z$, can get trapped between them forever (two sample configurations where this happens for $z=4, \tau = \frac{1}{4}, r_s=0.5$ and $z=4, \tau = \frac{2}{4}, r_s=1$ are shown in Fig.~\ref{fig:Trapping_States}). This means that we will get a non-zero value of $\eta$ for a relatively low density of agents $\rho$ for $r_s \leq min\{2 \tau, 1\}$ as can be verified from Fig.~\ref{fig:Inefficiency_Vs_Density}. Although for $r_s \leq min\{ 2 \tau, 1\}$ and at low densities, there are many configurations where the system eventually reaches the absorbing state, a handful of configurations exists where losing agents are trapped bounded by clusters of winning agents. At low densities, such configurations may only be realized over repeated random realizations. But such trapping states are guaranteed, even when the system size tends to infinity, and hence, theoretically, critical density tends towards zero for $r_s \leq min\{ 2 \tau, 1\}$.

For $r_s > min\{ 2 \tau, 1\}$, the system has a non-zero value of $\rho_c$. In this case, we can see that the system exhibits three distinct behaviors in inefficiency with respect to its variation with density. For low values of density below $\rho_c$, the system's inefficiency is zero. This is because, for low densities, the number of vacant sites is large, and with $r_s > min\{ 2 \tau, 1\}$, the competing agents can easily relocate and find sites whose neighborhoods are sparsely populated. The global inefficiency assumes a non-zero value at $\rho_c$ and continues to rise to finally peak near $\rho_p$ defined earlier (see Eqs.~\ref{eqn:PeakDensity_tau_lessThanHalf}, \ref{eqn:PeakDensity_tau_greaterThanHalf_a} \& \ref{eqn:PeakDensity_tau_greaterThanHalf_b}). We can see from Fig.~\ref{fig:Inefficiency_Vs_Density} that in this second range of density ($\rho_c<\rho\lesssim\rho_p$), the inefficiency is non-zero but is lower than that of a random configuration. There are still absorbing states available in this range of $\rho$, but even with $r_s >1$, the system is set to unending dynamics and never achieves the highest efficiency possible.

The difference in the observed behavior for variation of inefficiency with density for $r_s \leq min\{ 2 \tau, 1\}$ and $r_s> min\{ 2 \tau, 1\}$ is pronounced only when $\rho \lesssim \rho_p$. As the density increases above $\rho_p$ the inefficiency gets closer to that of a random configuration and shows a declining trend. The number of vacant sites is very small, and even a random arrangement of agents could distribute the few available vacant sites across the system. 

As $\rho \rightarrow 1$, most of the sites become occupied. The relevance of dynamics starts to fade at these densities, and the system's inefficiency remains equal to that of a random configuration. Hence, the very high-density range ($\rho \sim 1)$ is also categorized as the inefficient phase of the system, although the wastage of resources compared to the best possible arrangement is low here.  

\begin{figure}
\includegraphics[width=0.48\textwidth]{ 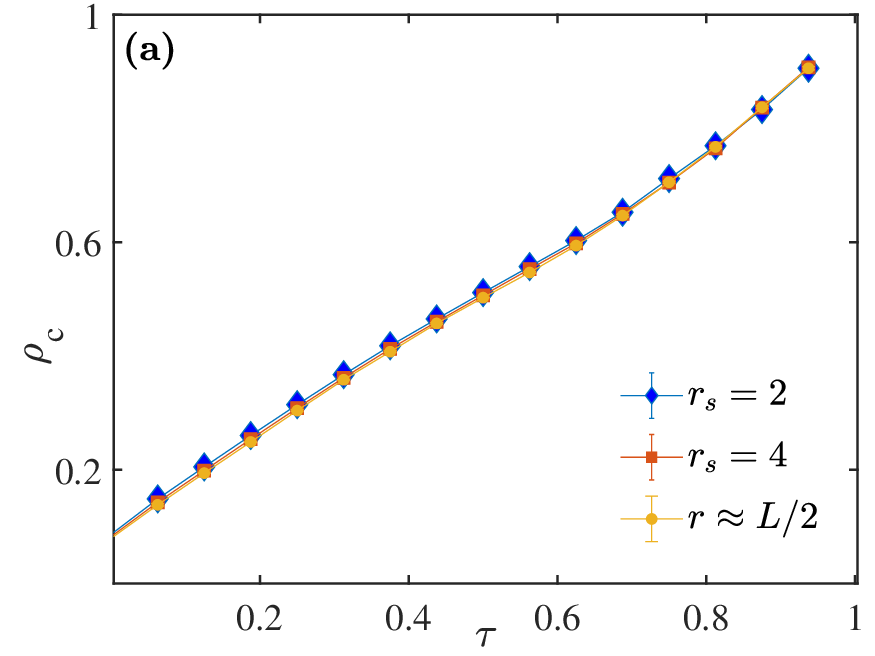}
\includegraphics[width=0.48\textwidth]{ 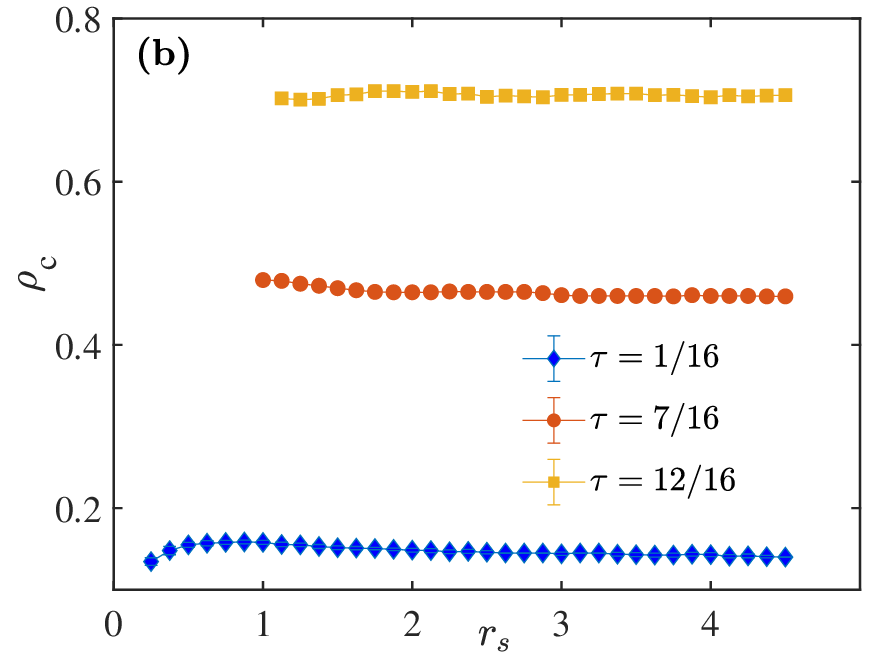}

\caption{Variation of critical density with \textbf{(a)} tolerance threshold and \textbf{(b)} standardized information radius for a $z=16$ system. The critical density ($\rho_c$) increases approximately linearly with the tolerance threshold and remains relatively constant as $r_s$ increases. Observations are averaged over $10^2$ trials.}  
\label{Critical_density}
\end{figure}

In Fig.~\ref{Critical_density}(a), we show the variation of $\rho_c$ with $\tau$ for $r_s> min\{ 2 \tau, 1\}$ for a $z = 16$ system. The critical density shows a linear growth with an increasing tolerance threshold for a constant value of $z$. Pearson's correlation coefficients, which measure the correlation between critical density and tolerance threshold, are approximately 0.999 for different $r_s$ values. The corresponding $p$-values are close to zero (approximately $10^{-19}$), indicating a robust linear correlation \cite{bibPearson}. This behavior of $\rho_c$ reflects how adaptive agents with larger tolerance thresholds can sustain systems of higher density at better efficiency. In Fig.~\ref{Critical_density}(b), we plot the variation of $\rho_c$ with $r_s$ for $r_s > min \{2\tau,1\}$. It is observed that, in general, $\rho_c$ remains constant as $r_s$ varies. 

\subsection{Information radius and its impact on the emergent behavior}

The variation in the information radius is significant only for the density range above $\rho_c$. For $\rho<\rho_c$, the system evolves into one of its absorbing states with zero inefficiency. Above $\rho_c$, we observe two diametrically opposite responses for inefficiency with respect to variation in the information radius. The peak sustainable density approximately demarcates the change in response.

\begin{figure}
\includegraphics[width=0.48\textwidth]{ 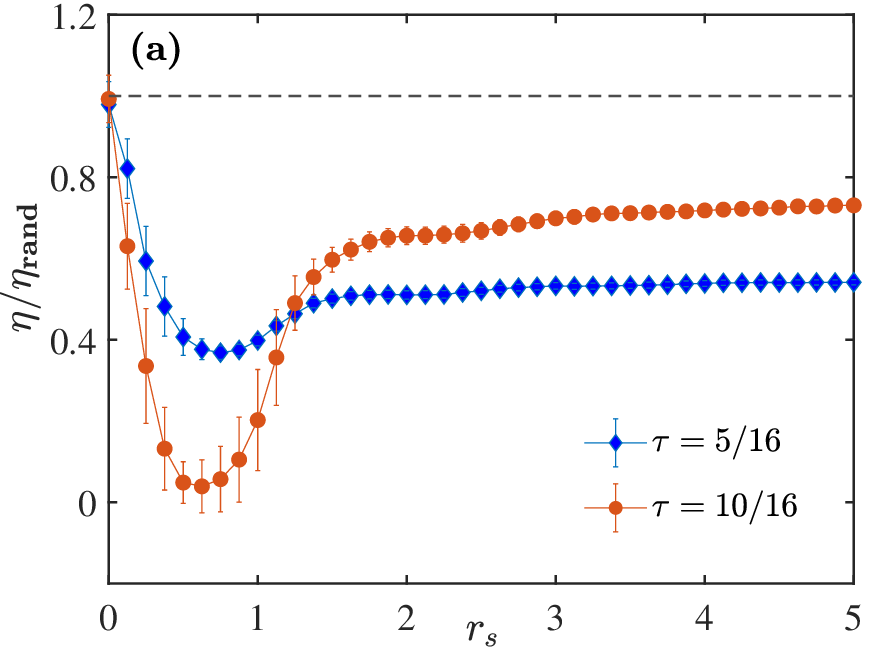}
\includegraphics[width=0.48\textwidth]{ 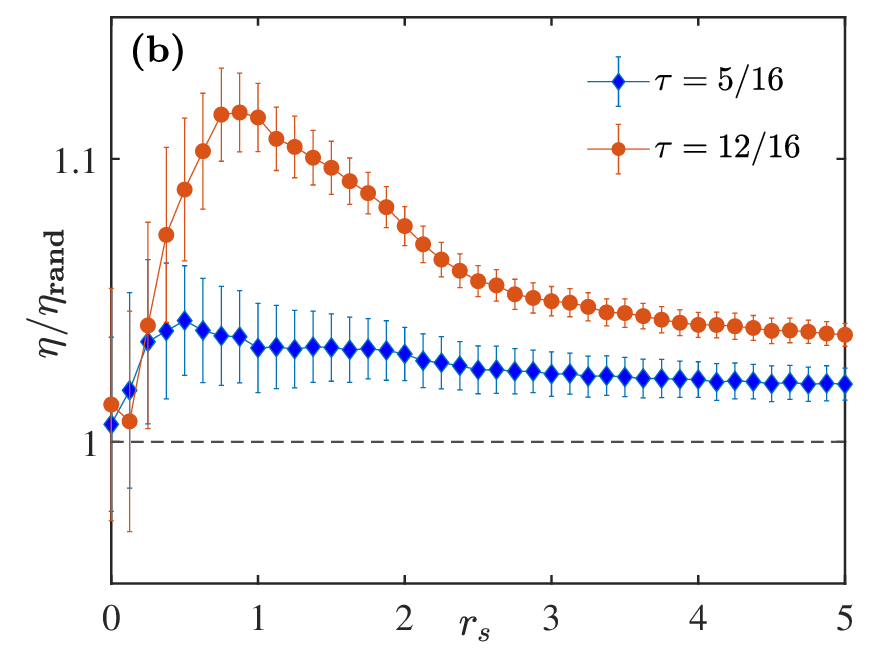}
\caption{The variation of inefficiency per agent (normalized to the corresponding random inefficiency) with standardized information radius for $z=16$ system ($L=201$) for \textbf{(a)} $\rho \lesssim \rho_p$ and \textbf{(b)} $\rho > \rho_p$. $Y-$axis is normalized with respect to $\eta_{rand}$ so that the dashed line represents the inefficiency corresponding to random configuration. For densities below $\rho_p$, the system achieves better coordination for comparatively lower ranges of $r_s$, and the inefficiency shows a minimum at an optimal value. For densities much above $\rho_p$, the system performs worse than random for all values of $r_s$, in stark contrast to the case with $\rho \lesssim \rho_p$.}  
\label{fig:Inefficiency_Vs_Information_Radius}
\end{figure}

Fig.~\ref{fig:Inefficiency_Vs_Information_Radius}(a) shows the variation of inefficiency with respect to the standardized information radius $r_s$ for a $z=16$ system at two different values of $\tau$ for $\rho_c < \rho \leq \rho_p$. Contrary to the notion that the more information, the better could be the agent's performance and hence the self-organization of the system, lower inefficiency is observed at lower information radii in this density range. As the information radius gets larger, the randomness in the decision-making and arrangement process hinders the efficient arrangement of agents in the system. High efficiency of the system requires ordering at a very local level, which essentially happens when the information radius roughly matches the neighborhood size. In these systems, an optimum value of $r_s$ exists at which the inefficiency is the least as evident from Fig.~\ref{fig:Inefficiency_Vs_Information_Radius}(a). 

\par The above result is in line with a few existing studies which show the irrelevance of high amount of input information in complex systems and how even highly detailed information can turn counter-productive for the system coordination and individual pay-offs \cite{bib2mandN, bib2mandN_2,bibMemoryRelevance,bibsv1, bibsv2,bibKushal1}. Our model exhibiting better adaptation at lower values of information radius below peak sustainable density emphasizes the fact that better information, whether in quality or quantity, need not always result in better outcomes.

As the system's density rises above $\rho_p$, the variation of inefficiency with respect to the standardized information radius shows an inversion of behavior. The system exhibits significant inefficiency in the lower information range and performs worse than a random distribution, as illustrated in Fig.~\ref{fig:Inefficiency_Vs_Information_Radius}(b). The highest inefficiency is observed at $r_s \sim 1$, paradoxically, which is the optimal information range for densities below the peak sustainable density. As the information radius increases further, the inefficiency of the system shows a slight decline. 

Another point to be noted is that the information radius ($r$) need not be larger than the neighborhood radius ($z/2$) for the system to self-organize. This is evident from the better global efficiency of the system observed even for $r_s < 1$ (Fig.~\ref{fig:Inefficiency_Vs_Information_Radius}(a)). In these systems, individual-level information and the corresponding behaviors can be thought to generate a \textit{collective intelligence} which in turn drives the system to higher levels of coordination and adaptation \cite{bibcollectiveintelligence, bibcollectiveminds}. 

\begin{figure*}
 \centering
 \includegraphics[width=0.99\textwidth]{ 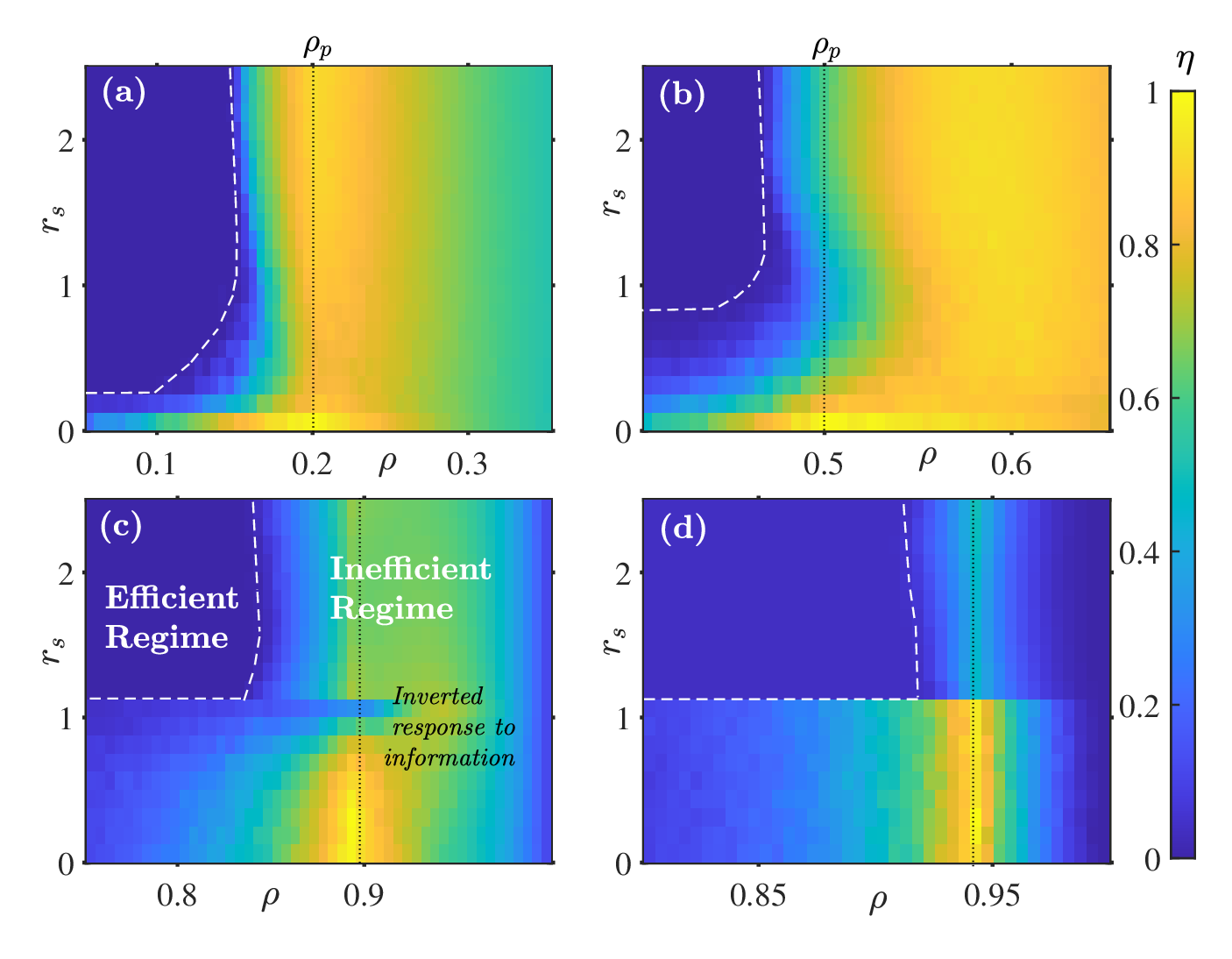} 
  \caption{Colormap in $\rho - r_s$ plane showing the variation of inefficiency per agent for a $z=16$ system with tolerance threshold values \textbf{(a)} $\tau=0$, \textbf{(b)} $\tau=\frac{7}{16}$, \textbf{(c)} $\tau=\frac{14}{16}$  and, \textbf{(d)} $\tau=\frac{15}{16}$. The transition from the efficient to the inefficient phase is demarcated horizontally by the critical density and vertically by the $r_s= min \{2\tau,1 \}$ line (shown in the figures using white dashed lines). The variation of inefficiency with information radius reverses its behavior at densities a bit above the peak sustainable density (shown in the figures using black dotted lines). The response of inefficiency to information shows features that are highly dependent on the tolerance threshold. As visible in the figures, the region of inverted response to information, observed above the peak sustainable density, gradually changes its shape and size and disappears. (Each data point is averaged over $10^2$ trials.)}  
\label{fig:Phase_diagrams_density-information}
\end{figure*}

Fig.~\ref{fig:Phase_diagrams_density-information} plots the phase diagram of the model in the $(\rho,r_s)$ plane and shows the impact of incremental changes in standardised information radius and density on the global inefficiency. The efficient phase is observed until $\rho_c$, up to which the system attains zero inefficiency. Above $\rho_c$, the system's inefficiency displays a highly dependent behavior on the input information and shows two contrasting responses to the input information ($r_s$) above and below $\rho_p$. In the efficient phase, the model consistently attains the absorbing state, while in the inefficient phase, a fluctuating steady state is obtained. 

The response of the system's inefficiency to the information radius depends on the tolerance threshold of the agents, as evident from Fig.~\ref{fig:Phase_diagrams_density-information}. For changes in the tolerance threshold, the response of inefficiency to the standardized information radius alters its behavior. The region of inverted response to information, observed above $\rho_p$, gradually changes appearance before vanishing for the highest tolerance threshold. 

\begin{figure*}
    \includegraphics[width=.48\linewidth]{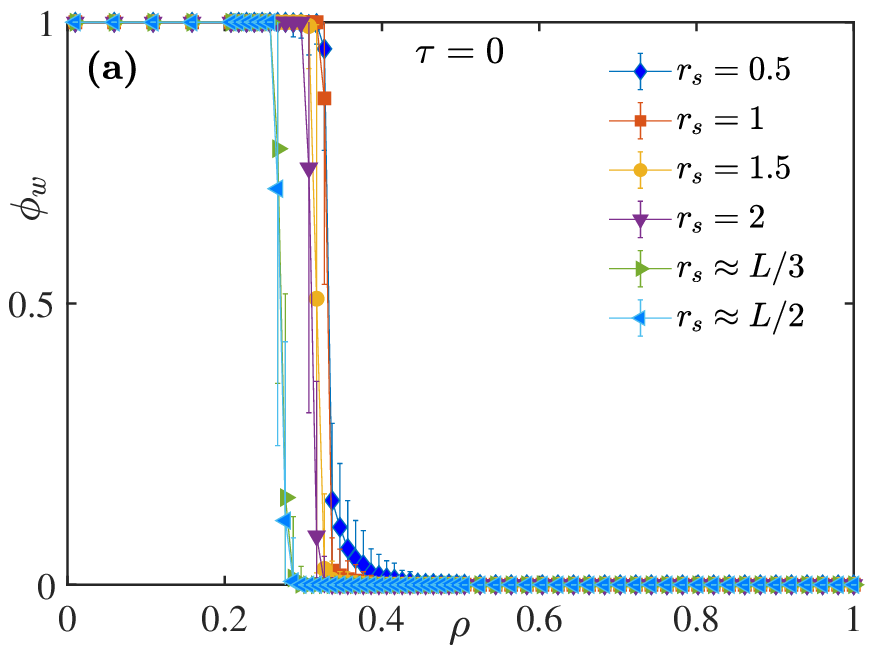}%
    \includegraphics[width=.48\linewidth]{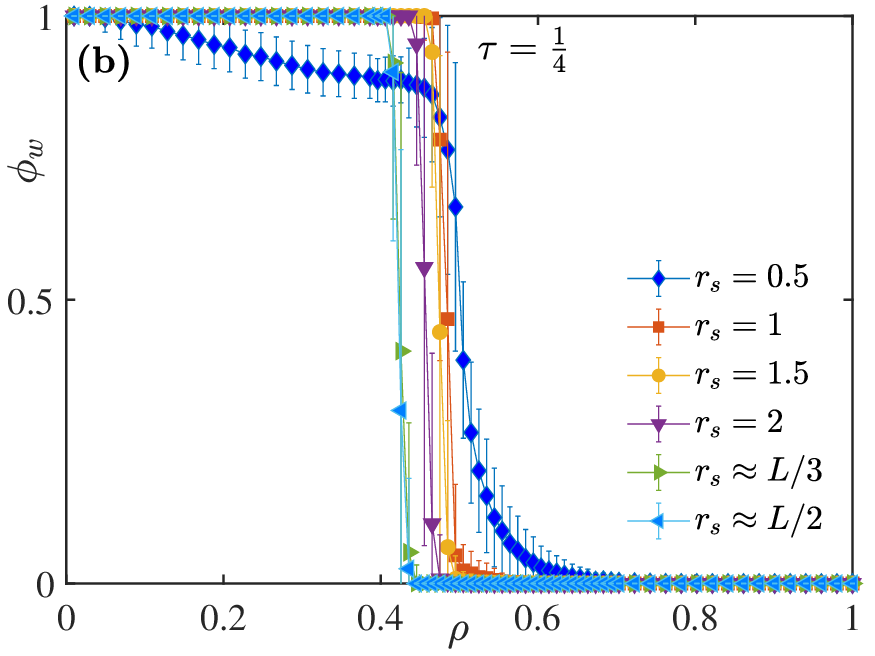}\\
    \includegraphics[width=.48\linewidth]{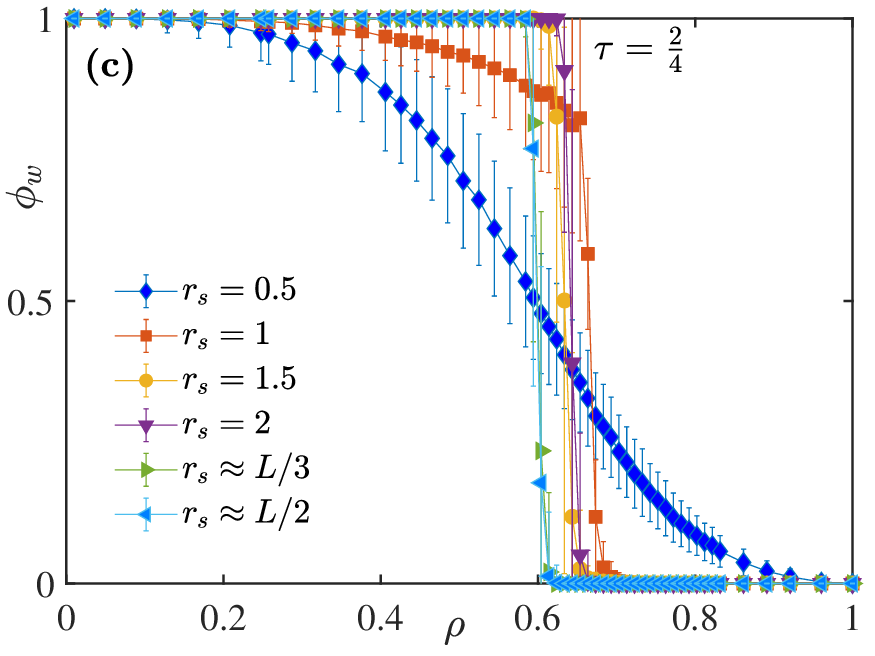}%
    \includegraphics[width=.48\linewidth]{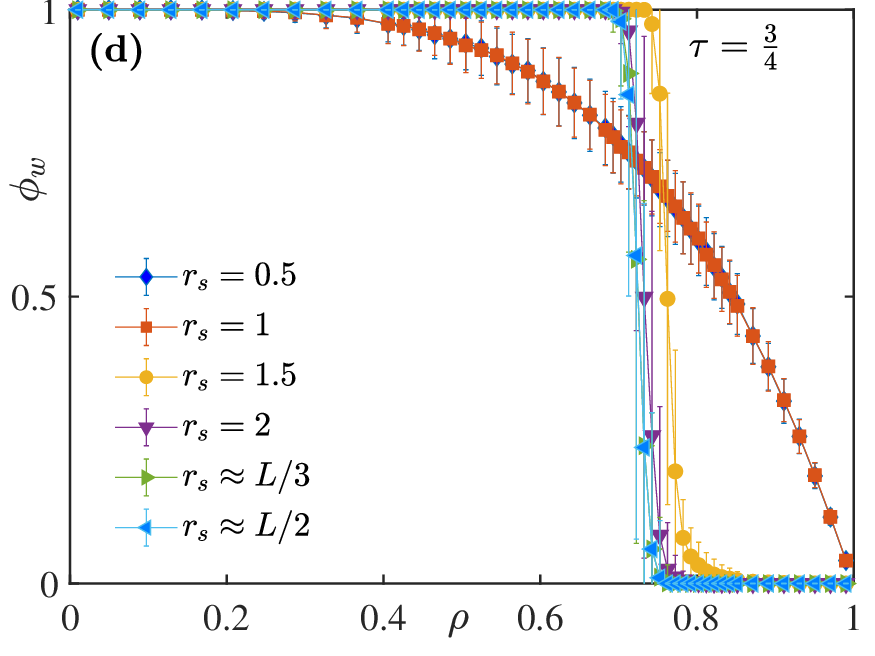}%
     \caption{Fraction of permanent winners Vs Density curves for $z=4$ system for various values of tolerance thresholds \textbf{(a)} $\tau=0$, \textbf{(b)} $\tau=\frac{1}{4}$, \textbf{(c)} $\tau=\frac{2}{4}$  and, \textbf{(d)} $\tau=\frac{3}{4}$. The simulations were performed on an $L=101$ system. The fraction of permanent winners ($\phi_w$) drops from 1 to 0 when the system transits from the efficient to the inefficient phase.}
        \label{fig:PermWinners_Vs_Density}
\end{figure*}

\subsection{Behaviour of fraction of frozen agents \texorpdfstring{($\phi$)}{phi} and Gini coefficient}

Agents who are permanent winners and remain in their positions form an interesting group of agents whose behavior can give insights into the internal dynamics of the system. Their fraction in the population ($\phi_w$) explains the critical density values. $\phi_w=1$ signifies the attainment of the absorbing state, and hence $\phi_w$ can be used as an order parameter to indicate the transition in the system. As observed in Fig.~\ref{fig:PermWinners_Vs_Density}, for $r_s> min\{ 2 \tau, 1\}$, the fraction of permanent winners ($\phi_w$) suddenly drops from 1 to 0  when the system transits from the efficient phase to the inefficient phase. 

As discussed in Section~\ref{Density_and_Inefficiency}, for $r_s \leq min\{\tau z,1\}$, there are states where losing agents are permanently bound between immobile winners. Therefore, the curves corresponding to $r_s \leq min\{\tau z,1\}$, exhibit a decrease from 1 starting from $\rho \approx 0$. When the system remains completely static, for $\tau = \frac{z-1}{z}$ \& $r_s<1$, the fractions of the permanent winners are the same as that of a random configuration. In this case, the fraction of the permanent winners is given by

\begin{equation}
    \phi_w=1-\rho^{\tau z+1}
\end{equation}

In Fig.~\ref{fig:PermWinners_Vs_Density}(d), the cases for $r_s<1$ correspond to this scenario. 

\begin{figure}[t]
\includegraphics[width=0.48\textwidth]{ 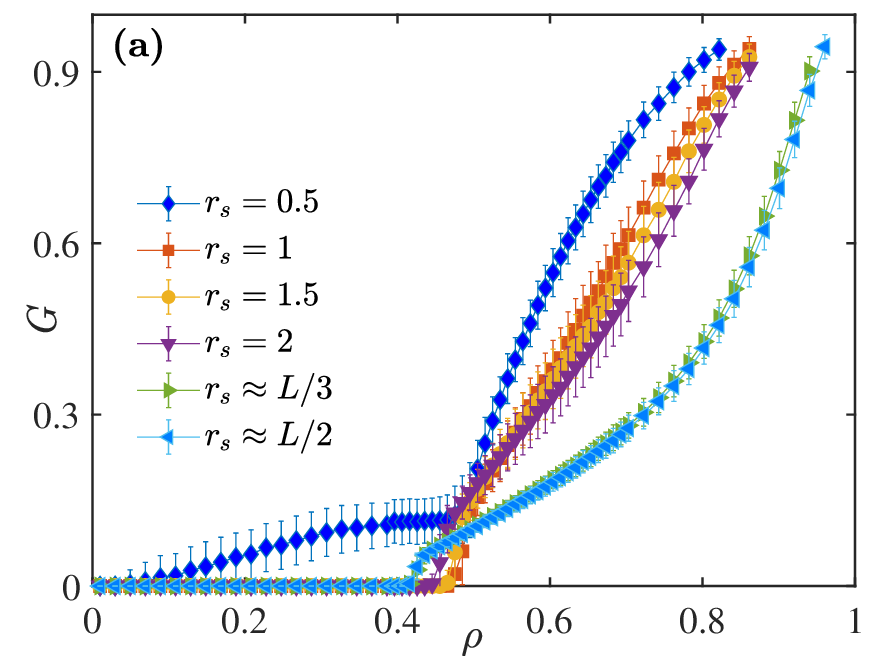}
\includegraphics[width=0.48\textwidth]{ 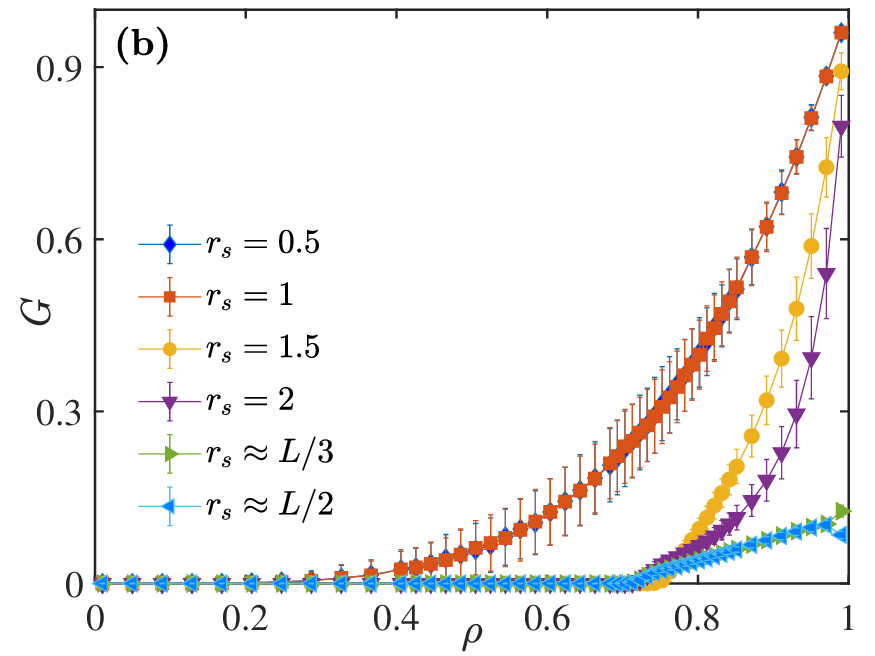}
\caption{Variation of Gini coefficient $G$ with respect to population density $\rho$ for a system with \textbf{(a)} $z=4; \tau=0.25$ and \textbf{(b)} $z=4; \tau=0.75$ for different information radii $r_s$. For $r_s > min\{2 \tau,1\}$, $G$ turns non-zero at the critical density ($\rho_c$).}  
\label{fig: G_Vs_Density}
\end{figure}

 The inequality in agents' wealth quantified using the Gini coefficient $(G)$ exhibits interesting characteristics when studied as a function of density and information radius. Fig.~\ref{fig: G_Vs_Density} shows the variation of $G$ with density for different standardized information radii $(r_s)$. The Gini coefficient $(G)$ changes from zero to a non-zero value at $\rho_c$. Thus, $\rho_c$ marks the onset of not only inefficiency, but also inequality in a dynamic system. 
 
\begin{figure}[t]
\includegraphics[width=0.5\textwidth]{ 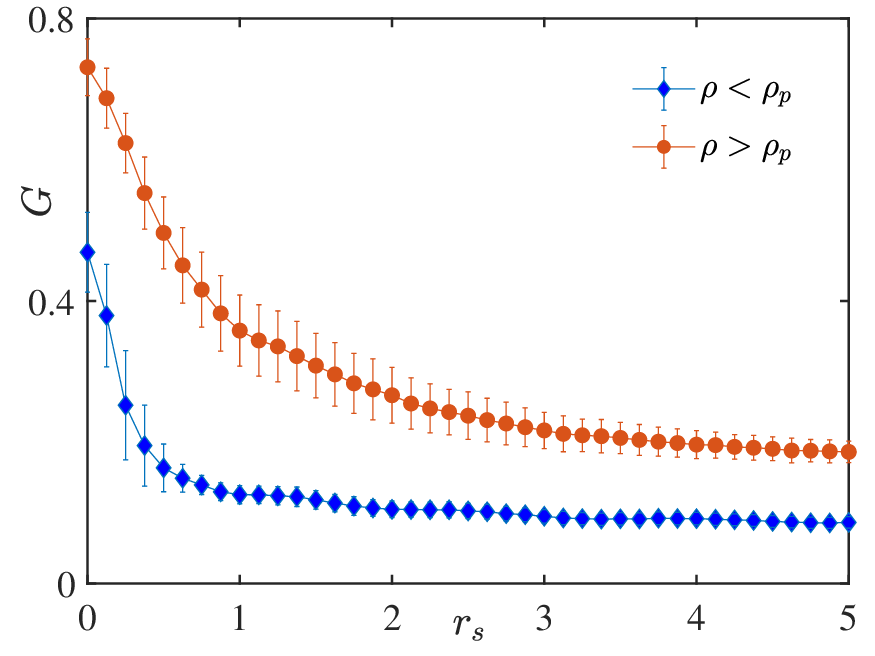}%
\caption{Variation of Gini Coefficient $G$ with standardized information radius $r_s$ for a $z=16; \tau = \frac{12}{16}$ system below and above the peak sustainable density $\rho_p$. The pay-off inequality decreases with an increase in $r_s$ in both cases. Note that $r_s = 0$ corresponds to the case of random distribution of agents.}  
\label{fig: Gini_Vs_InformationRadius}
\end{figure}

 The Gini coefficient decreases with an increase in the agents' information radius, as observed in Fig.~\ref{fig: Gini_Vs_InformationRadius}. The observation holds for most density ranges, except for a narrow range immediately following $\rho_c$ visible in Fig.~\ref{fig: G_Vs_Density}. Thus, even when the role of information in reducing inefficiency remains variable, it is safe to generalize that a higher information radius almost always guarantees better equality in the system. 

\section{Conclusion and discussion}\label{Conclusion}

The local interactions and preferences of individual agents determine the large-scale statistical properties in many complex systems. In the past, researchers have looked into in detail how different agent preferences like alignment with neighbors (as in voter models \cite{VoterModel_Clifford_1973, Holley_Liggett_1975, VoterModel_Coarsening_Persistence, Liggett_book_1999, Liggett_Stochastic_Models}), anti-alignment with neighbors (as in contrarian voter models\cite{Gimenez2022_Anti-alignment_Voter}), affinity to be among a particular type of agents (as in Schelling's model of racial segregation \cite{schelling_Micromotives_Macrobehavior, dynamics_schelling_model, Schelling_phase_diagram, StatPhysics_Schelling_model}), etc. affect global properties in a spatial setting. Strategic interaction between the neighboring agents is also considered extensively (As in spatial Prisoner's Dilemma \cite{Nowak_4_SpatialGames_Cooperation, Spatial_PD_adaptive_migration, moyano2009_Cooperation_SpatialPD}, Stag Hunt \cite{ bibstaghunt, sicardi2009RandomMobility_SpatialStructure_PD_SH_SD}, Hawk-dove \cite{bibbibHawkesDovesSpace} and Snow Drift Games \cite{bibsnowdrift, Spatial_Snowdrift_myopicAgents, Network_ChickenGame}). 

In this work, motivated by examples of complex systems involving competition and sharing of physical space, we consider a different type of agent preference—local crowd avoidance—in a spatial setting. In the model, competing agents strive for advantageous locations by avoiding the local crowd, trying to keep the local population density in their neighborhood below their tolerance threshold by relocation. We study the resultant emergent behavior on a 1D lattice system. The crowd avoidance model we define thus lies in the broad class of spatial models that investigate how a group of agents respond and form collective action and emergence based on spatial or neighborhood preferences. 

On a preliminary level, there is some similarity with Schelling's model of segregation in that both models trace patterns of emergence based on the requirement of having (or not having) a minimum fraction of a kind of agents in the neighborhood. However, the present model stands apart in that it focuses on understanding spatial crowd-avoidance behavior. The micro-mechanisms used in defining the rules and settings of the problem reflect this. For example, there is only one kind of agent, and empty sites that have neither preferences nor actions are actually the resource for which they compete. Furthermore, on the macroscale, unlike the Schelling's model which focuses on the emergence of segregation, we investigate the competitive allocation of spatial resources.

The problem of spatial competition is often addressed in the literature in other contexts as well. Traditional economic models, such as the Hotelling model and its several variants, examine how firms optimize their location and pricing of products to capture market share along a continuous line \cite{HotellingModel_Drezner2024,Competition_Hotelling1929}. In contrast to such settings, our model focuses on competition for space in a different context, where multiple agents try to optimize their comfort within a locally defined neighborhood with a tolerance threshold for local crowding. Specifically, there is no explicit pricing aspect and the agents may accommodate others in their neighborhood depending upon their tolerance threshold. Unlike the Hotelling model, which typically involves competition by the agents (usually two) across the entire space, the agents in our problem make decisions based only on local information. We may say that in both models, the agents self-organize across the space based on their optimization goals, although no explicit dynamics is usually considered in the Hotelling model framework. Interestingly, despite these differences, it seems that in some specific scenarios, both models can evolve toward similar equilibrium states. For example, when the neighborhood size is $\sim L/N$ and the tolerance threshold is {\it zero}, agents in our problem repel each other and spread uniformly across the lattice. This behavior mirrors the Nash equilibrium of the Hotelling model on a ring with $N$ agents, where the agents are positioned equidistantly \cite{HotellingCircular_Gong2016}.

There are numerous models in the literature that study the emergent behavior due to particle dynamics on a lattice, such as exclusion processes, lattice gas models,  zero-range process etc~\cite{ASEP_DERRIDA1998, HPP1973_latticeGas, ZRP_SPITZER1970}. These models share a commonality with ours in that they are based on the movement of entities on a lattice, with fundamental mechanisms built around neighborhood interactions. However, our model diverges from these in its core objective, dynamics, and emergent behavior. For example, in the lattice gas models, primarily used for fluid dynamics and cell movement, space serves merely as a backdrop for particle or agent movement~\cite{Bio_LGCA_2021Detsch,CellularAutomata_Deutsch2005}. In contrast, our model focuses on spatial resource utilization and how it is affected by characteristics like density and range of relocation of agents. Agents within our model interact with their environment not simply by moving through it, but by dynamically responding to local crowding in space
in an effort to optimize their usage of available space. This fundamental difference shapes the mechanisms: in our model, agents dynamically rearrange in response to local crowding, driven by an active goal-oriented behavior. The emergent behaviors also differ, as we measure how efficiently agents distribute themselves in space, emphasizing spatial resource competition without external control.

In the model, first we analytically calculated the densest possible arrangement of agents for a given value of neighborhood size and tolerance threshold and arrive at the expressions for the highest density (termed the peak sustainable density) at which the system can be packed, ensuring that all the agents are winners. Then, we simulated the model and analyzed the emergent behavior. The dependence of its macroscopic features - inefficiency, inequality, and the fraction of frozen agents in the system - on the population density and the information accessibility of the agents is studied. 

We find that density is an important determinant of the emergence of the system. A dynamic system manifests the emergence of order and maintains zero inefficiency up to the critical density. If the density is below $\rho_c$, the system evolves into one of its absorbing states. Above a higher density called the peak sustainable density ($\rho_p$), there are guaranteed losers in the system. Between $\rho_c$ and $\rho_p$, inefficiency is non-zero but remains lower than that of a random arrangement signaling emergent coordination between the agents. When the system's density is above $\rho_p$, the global inefficiency of the system starts to decline and approaches that of a randomly arranged configuration of agents.
 
 Interestingly, the global inefficiency shows a non-monotonic dependence on the information accessibility of agents and displays two opposite characters above and below $\rho_p$. Below $\rho_p$, highly efficient system organization and agent adaptation are observed for smaller standardized information radius revealing how smaller information inputs may favor better system efficiency. There also exists an optimal information radius at which the inefficiency is minimum. Above the peak sustainable density, the system reverses its behavior in response to variations in the information radius.  In this regime, higher inefficiency occurs at lower information radii. The varying impact of information inputs on the global system behavior suggests that the prudent choice of information cannot be based on either the quantity or the quality of the information but depends on the parameters of the system and the resultant complex interaction. 

In contrast, the variation of the Gini coefficient, which measures the inequality between agents, with respect to information radius displays a consistent behavior for most density ranges. High inequality is observed at lower information radii, and the inequality falls off rapidly with an increase in the information radius for most of the density values except for a narrow density range near $\rho_c$. Therefore, although the role of information in reducing inequality is contingent on other conditions, its role in bringing down inequality is significant. 

Our study opens multiple avenues for further investigations into the unexplored realms of spatial crowd avoidance behavior and the resulting dynamics. An automatic extension of the problem is to look into the features of higher-dimensional systems and spaces with different structures. Attempts can be made to solve the system analytically, with the relocation of agents restricted to only the neighboring sites of an agent which could be more amenable to analytical treatments. The current model employs agents obeying a static and homogeneous rule with a sequential update scheme. This can be generalized by introducing adaptive and heterogeneous behaviors and updating rules. An interesting area where the heterogeneity of agents can bring useful insights is in addressing questions regarding the impact of information asymmetry on outcomes at both individual and global levels when some agents have access to more information. This will help us to understand the role of information in complex systems in a more comprehensive manner. The quantifiable nature of individual-level and collective information in our model provides ample scope for investigating the emergence of collective intelligence in the context of spatial resource sharing in future studies. 

\bibliography{Bib}
\appendix
\section{General expressions for the peak sustainable density} \label{AppendixA}
We formulate general expressions for the peak sustainable density in terms of the neighborhood size $z$ and tolerance threshold $\tau$ of the agents. In formulating these expressions, we limit our search to configurations with clusters of equal size.  Following the method described in Section \ref{Peak_Sust_Density}, we compose configurations of winning agents with cluster sizes varying from the maximum possible value of $\tau z +1$ to the minimum possible value of 1. The highest density among these configurations is identified as the peak sustainable density.

We have seen that the neighborhoods of the agents in a cluster begin to expand into neighboring clusters and cluster gaps when the cluster size is gradually decreased from $\tau z +1$. As stated in section \ref{Peak_Sust_Density}, if removing one agent from a cluster results in all the other agents gaining additional vacant sites in their neighborhood then the cluster gap size too can be reduced accordingly. On the other hand, the cluster gap size cannot be reduced if removing one agent results in the neighborhood of any of the remaining agents in the cluster extends into an occupied site as a result. Using these facts, in the following, we formulate the general expressions for the peak sustainable density. We treat the cases $\tau < \frac{1}{2}$ and $\tau \geq \frac{1}{2}$ separately because they differ in the access of the agents in the largest possible cluster to the nearest cluster gaps.

\subsection{For tolerance threshold \texorpdfstring{$\tau < \frac{1}{2}$}{tau < 1/2}}

\begin{figure}
     \includegraphics[width=.48\linewidth]{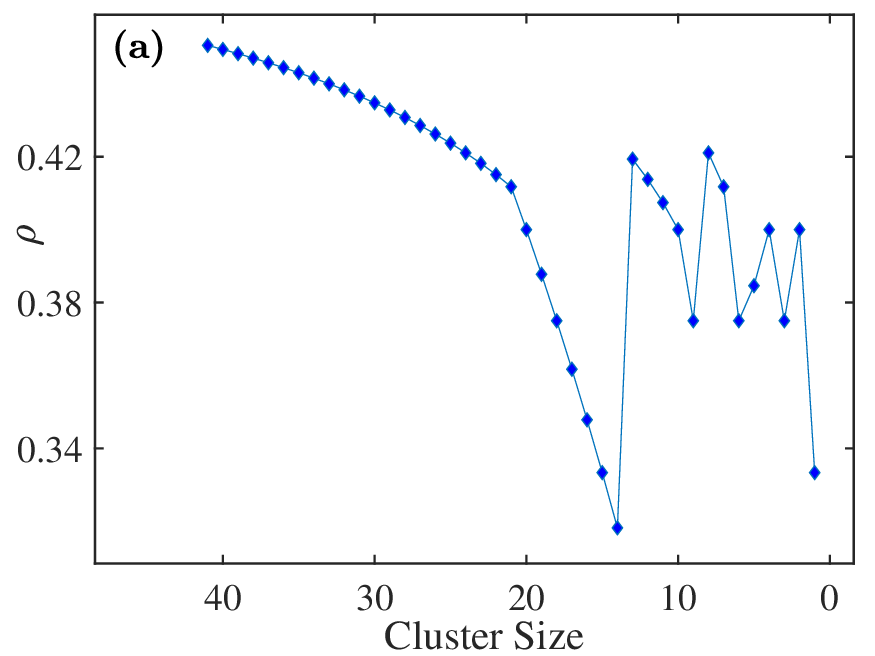}
    \includegraphics[width=.48\linewidth]{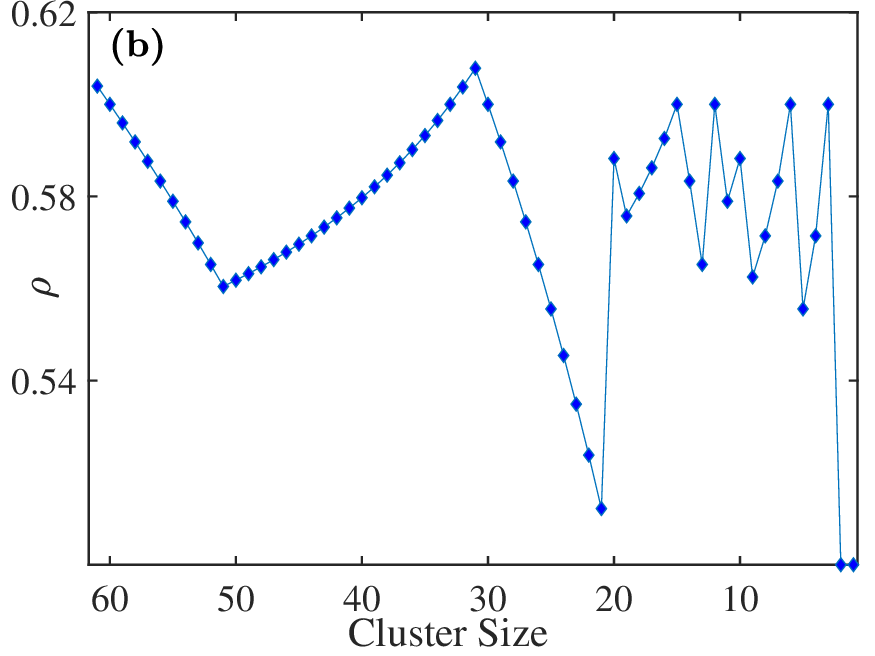}%
    \caption{Variation of density with cluster size of a winning configuration for $z=100$ for \textbf{(a)} $\tau= \frac{40}{100} (\tau <\frac{1}{2})$ and \textbf{(b)} $\tau= \frac{60}{100} (\tau \geq \frac{1}{2})$. The highest value of density is the peak sustainable density.}
        \label{fig:Packing Density_Vs_Cluster Size}
\end{figure}

In this case, the largest possible cluster size is ($\tau z +1$) with a cluster gap of size $\frac{z}{2}$. The subsequent lower-sized clusters of winning agents with maximum density can be created by successively removing the agents, keeping the cluster gap size at an essential minimum while doing so. It can be seen that initially, for the removal of a few agents from the largest possible cluster, the cluster gap size can be reduced along with the cluster size. Fig.~\ref{fig:Packing Density_Vs_Cluster Size}(a) shows the variation of density of agents with respect to variation in cluster size for such a sequence of configurations for a particular value of $z$ and $\tau$.

Let $i$ denotes the number of agents removed from the biggest possible winning cluster described in the manner mentioned above. The expression for the density of agents for the series of configurations generated for increasing values of $i$ can be written as,

\begin{equation}
\rho= \frac{\tau z+1-i}{(\tau z+1-i)+ (z/2 -i)} \text{ ; where }  i=0,1,2 \ldots i_{max}
    \label{eqn: Density_tau_lessThanHalf_i}
\end{equation}

Here, $i_{max}$ is the maximum number of agents that can be removed from the biggest possible winning cluster, such that the cluster gap can also be reduced concurrently. Thus, $i_{max}$ corresponds to the case when the neighborhoods of all the agents cover the two nearest cluster gaps on either side and any further removal of agents results in the neighbourhood of at least one of the agents in the cluster extending into the next nearest cluster (i.e., to an occupied site). At this point, the cluster gap size will be ($\lceil \frac{z-\tau z}{2}\rceil$), and the cluster will be of size ($\tau z +1-i_{max}$).  Therefore, $i_{max}$ can be expressed as,

$$i_{max}=\frac{z}{2}- \lceil \frac{z-\tau z}{2}\rceil $$

It is easy to see that the highest value for $\rho$ from Eq. \ref{eqn: Density_tau_lessThanHalf_i} is obtained for $i=0$. This is because, for $\tau < \frac{1}{2}$, peak sustainable density ($\rho_p)$ is less than or equal to $\frac{1}{2}$, and subtracting one from the numerator and two from the denominator of a fraction that is less than $\frac{1}{2}$ always yields a smaller fraction.

When the cluster size is reduced even further, the cluster gap size can not be reduced correspondingly since agents require at least $z - \tau z $ vacancies in their neighborhood to win. This vacancy requirement should be met by the two cluster gaps of fixed size $ \lceil \frac{z-\tau z}{2} \rceil$ on both sides. Hence, packing density decreases at a faster rate when the cluster size is reduced beyond $i_{max}$ (see Fig.~\ref{fig:Packing Density_Vs_Cluster Size}(a)). (If $z-\tau z$ is an odd number, then configurations with cluster gaps of sizes  $\lceil \frac{z-\tau z}{2} \rceil$ and $\lfloor \frac{z-\tau z}{2} \rfloor$ asymmetrically distributed on both sides are possible. However, it should be noted that such configurations have a lower packing density compared to configurations with larger cluster sizes).

This decreasing trend of the packing density will continue until the neighborhood of all the agents in the cluster starts to have access to the cluster gaps which are further away (third, fourth, etc.) at which the packing density will show momentary spiking as visible in Fig.~\ref{fig:Packing Density_Vs_Cluster Size}(a).

\subsection{For tolerance threshold \texorpdfstring{($\tau \geq \frac{1}{2}$)}{tau >= 1/2}}

 Fig.~\ref{fig:Packing Density_Vs_Cluster Size}(b) shows the variation in the density of configurations where all agents win when the cluster size is decremented from the maximum to the minimum possible value for a sample system with $\tau \geq \frac{1}{2}$. In this case, the largest cluster size possible with all agents winning is ($\tau z +1$), but with a cluster gap of size ($z-\tau z$). Here, unlike in the earlier case, the agents at the edge of the biggest possible winning cluster have access to only one cluster gap. Therefore, the cluster gap size cannot be reduced initially, when decrementing the cluster size from the largest possible value, as all agents must remain winning.

 Let $i$ denote the number of agents removed from the largest possible cluster without reducing the cluster gap size as described above. Then, the densities of the configurations realized by decrementing the cluster size can be written as follows:

\begin{align}
\begin{split}
\rho & = \frac{\tau z+1-i}{(\tau z+1-i)+ (z -\tau z)} \\
     & = \frac{\tau z+1-i}{z+1-i}  \text{ ; where }  i=0,1,2 \ldots i_{max}
    \label{eqn:Density_tau_GreaterThanHalf_i}
\end{split}
\end{align}

 Here, $i_{max}$ represents the maximum number of agents that can be removed from the largest possible winning cluster while maintaining a fixed minimum cluster gap size of $z-\tau z$. This also pertains to the case when the cluster is of the minimum size, with the agent at the edge accessing only one nearest cluster. Consequently, at this juncture, the cluster size ($\tau z +1-i_{max}$) is equal to ($\frac{z}{2}+1$). Therefore, $i_{max}$ can be represented as

\begin{equation}
   i_{max}=(\tau z +1) -(\frac{z}{2}+1)
\end{equation}

It is easily seen that the highest value among the set of possible values for different $i$ given by Eq. \ref{eqn:Density_tau_GreaterThanHalf_i} is for $i=0$, given by,

\begin{equation}
\rho  =  \frac{\tau z+1}{z+1}
    \label{eqn:Density_for_large_tau_2}
\end{equation}

 When the cluster size is further reduced from ($\frac{z}{2} + 1$), the neighborhoods of all agents in the cluster begin to expand to the second nearest cluster gap. In that case, the cluster gap size can also be reduced along with the reduction in the cluster size. Hence, the expression for the density of these configurations when the agents have access to the two nearest cluster gaps can be written as,

\begin{eqnarray}
\rho =
    \frac{\frac{z}{2}-j}{(\frac{z}{2}-j)+(z-\tau z- (j+1))} ;\\
    \text{ where }  j=0,1,2 \ldots j_{max}
    \label{eqn:Density_for_large_tau_3}
\end{eqnarray}

The first increasing trend of packing density when the cluster size is reduced, as seen in Fig.~\ref{fig:Packing Density_Vs_Cluster Size}(b), happens when the neighbourhood of all the agents in the cluster gain access to the second nearest cluster gap and thus follows Eq.~\ref{eqn:Density_for_large_tau_3}. This increasing trend in packing density will continue only until the neighborhoods of all the agents in the cluster entirely cover the two nearest cluster gaps, i.e. when $b= \lceil \frac{z-\tau z}{2} \rceil$. Thereafter, the packing density again starts to decrease. Hence, in Eq.~\ref{eqn:Density_for_large_tau_3}, the upper limit of $j$ can be determined using the condition,

\begin{equation}
(z-\tau z-(j+1)) \geq \frac{z-\tau z}{2}
\end{equation}

or,
\begin{equation}
j \leq \frac{z-\tau z}{2}-1
\end{equation}

which implies,
\begin{equation}
j_{max} = \lfloor \frac{z-\tau z}{2} \rfloor -1
\end{equation}

The highest value for Eq.~\ref{eqn:Density_for_large_tau_3} is obtained when $j=j_{max}$, since for $\tau \geq \frac{1}{2}$, peak sustainable density ($\rho_p$) is greater than $\frac{1}{2}$, which is given by,

\begin{equation}
\rho =
    \frac{\frac{z}{2}-j_{max}}{(\frac{z}{2}-j_{max})+ \lceil \frac{z-\tau z}{2} \rceil }
\label{eqn:Density_for_large_tau_4}
\end{equation}

It can be seen that the configurations generated with cluster sizes smaller than $(\frac{z}{2}- j_{max})$ have a packing density less than or equal to the case when the neighborhoods of all the agents in the cluster span the two nearest cluster gaps. This constraint is due to the requirement of having $(z- \tau z)$ vacancies in the neighborhood of each agent. The local maxima in the packing density following the initial two peaks, as seen in Fig.~\ref{fig:Packing Density_Vs_Cluster Size}(b), are attained when all agents in the cluster access the subsequent cluster gaps entirely.

 Thus, the peak sustainable density ($\rho_p$) for $\tau \geq \frac{1}{2}$ is the largest value among the set of fractions obtained from equations \ref{eqn:Density_for_large_tau_2} and \ref{eqn:Density_for_large_tau_4}. i.e., the densest packing is either for the one with the largest possible cluster size with all agents winning or for the one with a maximal cluster size with the neighborhood of all the agents in the cluster spanning the two nearest cluster gaps entirely. Note that numerically there is not much difference between the values given by Eqs.~\ref{eqn:Density_for_large_tau_2} and \ref{eqn:Density_for_large_tau_4} in the parameter range of $\tau$ we are interested in.

\end{document}